\documentclass[12pt]{article}
\usepackage{epsfig}
\usepackage{graphicx}
\usepackage{amsmath}

\begin{document}

 \newcommand{\beq}{\begin{equation}}
\newcommand{\eeq}{\end{equation}}
\newcommand{\bea}{\begin{eqnarray}}
\newcommand{\eea}{\end{eqnarray}}
\newcommand{\beqn}{\begin{eqnarray}}
\newcommand{\eeqn}{\end{eqnarray}}
\newcommand{\beas}{\begin{eqnarray*}}
\newcommand{\eeas}{\end{eqnarray*}}
\newcommand{\defi}{\stackrel{\rm def}{=}}
\newcommand{\non}{\nonumber}
\newcommand{\qt}{\tilde q}
\newcommand{\m}{\tilde m}
\newcommand{\trho}{\tilde{\rho}}
\newcommand{\tn}{\tilde{n}}
\newcommand{\tN}{\tilde N}
\newcommand{\pt}{\partial}

\newcommand{\gsim}{\lower.7ex\hbox{$
\;\stackrel{\textstyle>}{\sim}\;$}}
\newcommand{\lsim}{\lower.7ex\hbox{$
\;\stackrel{\textstyle<}{\sim}\;$}}


\def\Tr{ \hbox{\rm Tr}}
\def\const{\hbox {\rm const.}}
\def\o{\over}
\def\im{\hbox{\rm Im}}
\def\re{\hbox{\rm Re}}
\def\bra{\left\langle}
\def\ket{\right\rangle}
\def\Arg{\hbox {\rm Arg}}
\def\Re{\hbox {\rm Re}}
\def\Im{\hbox {\rm Im}}
\def\diag{\hbox{\rm diag}}
\def\d{\partial}

\def\Re{{\rm Re}}
\def\rank{{\rm rank}}
\def\2{{1\over 2}}
\def\ntwo{${\mathcal N}=2\;$}
\def\nfour{${\mathcal N}=4\;$}
\def\none{${\mathcal N}=1\;$}
\def\ntwot{${\mathcal N}=(2,2)\;$}
\def\ntwoo{${\mathcal N}=(0,2)\;$}
\def\x{\stackrel{\otimes}{,}}

\def\stackreb#1#2{\mathrel{\mathop{#2}\limits_{#1}}}

\newcommand{\vp}{\varphi}
\newcommand{\ca}{{\mathcal A}}
\newcommand{\cw}{{\mathcal W}}
\newcommand{\ve}{\varepsilon}
\newcommand{\cpn}{CP$(N-1)\;$}
\renewcommand{\theequation}{\thesection.\arabic{equation}}

\setcounter{footnote}0

\vfill

\begin{titlepage}

\begin{flushright}
FTPI-MINN-10/13, UMN-TH-2904/10\\
May 28, 2010
\end{flushright}

\begin{center}
{  \Large \bf  Moduli Space Potentials for  Heterotic\\[2mm]
 non-Abelian Flux Tubes: Weak Deformation
 }

\vspace{4mm}

 {\large
 \bf    M.~Shifman$^{\,a}$ and \bf A.~Yung$^{\,\,a,b}$}
\end {center}

\begin{center}


$^a${\it  William I. Fine Theoretical Physics Institute,
University of Minnesota,
Minneapolis, MN 55455, USA}\\
$^{b}${\it Petersburg Nuclear Physics Institute, Gatchina, St. Petersburg
188300, Russia
}
\end{center}

\begin{center}
{\large\bf Abstract}
\end{center}

We consider \ntwo supersymmetric QCD with the U($N$) gauge group (with no Fayet--Iliopoulos term)
and  $N_f$ flavors of massive quarks deformed by the mass term $\mu$
for the adjoint matter, $\cw=\mu\ca^2$, assuming that
$N\leq N_f<2N$. This deformation breaks \ntwo supersymmetry down to
\none\!.
This theory supports non-Abelian flux tubes (strings) 
which are stabilized by $\cw$.
They are referred to as $F$-term stabilized strings. We focus on the studies of such strings
in the   vacuum in which $N$ squarks condense, at small $\mu$,
so that the $Z_N$ strings preserve, in a sense, their BPS nature. We calculate string
tensions both in the classical and quantum regimes. Then we translate
our results for the tensions in terms of the effective low-energy weighted CP$(N_f-1)$ model on
the string world sheet. The bulk $\mu$-deformation makes this theory
\ntwoo supersymmetric heterotic weighted CP$(N_f-1)$ model in two dimensions.
We find the   deformation potential on the world sheet.
This significantly expands the class of the heterotically deformed CP models emerging on the string world sheet
compared to that suggested by Edalati and Tong.
Among other things,
we show that nonperturbative quantum effects in the bulk theory are exactly reproduced by the
quantum effects in the world-sheet theory.

\vspace{2cm}

\end{titlepage}



\newpage

\section {Introduction}
\label{intro}
\setcounter{equation}{0}

In this paper we will report on further developments in non-Abelian strings, a construction 
which emerged recently~\cite{HT1,ABEKY,SYmon,HT2} (for detailed reviews see  \cite{Trev,Jrev,SYrev,Trev2}).
Originally the non-Abelian strings were discovered \cite{HT1,ABEKY,SYmon,HT2} in \ntwo supersymmetric QCD with 
the U($N$) gauge group
and $N_f=N$ quark multiplets and the Fayet--Iliopoulos (FI) $D$-term \cite{FI}.
The role of the FI term is to trigger the quark condensation and 
provide stabilization for the BPS-saturated flux-tube solitons. 
The BPS nature of the flux tubes obtained in this way guarantees \ntwot supersymmetry
on the string world sheet.
The next step in this program was breaking \ntwo supersymmetry down to \none by virtue of the
$\mu\ca^2$ superpotential \cite{SYnone,Edalati,SYhet}. As a result of this deformation of the original \ntwo bulk theory,
supersymmetry on the string world sheet reduces (classically) from \ntwot down to \ntwoo \cite{Edalati,SYhet}.
This happens because a superpotential $\omega \sigma^2$ is generated on the world sheet.
The parameters $\mu$ and $\omega$ are related by a proportionality formula,\footnote{The proportionality  formula obtained in \cite{SYhet} is only valid in the limit of small $\mu$.} while the functional dependence of the superpotentials
in the bulk and on the world sheet is the same -- quadratic -- as was suggested in \cite{Edalati} and confirmed in
\cite{SYhet} by a direct calculation. Taking account of quantum effects on the
string world sheet one observes \cite{SYhet,SYhetN} spontaneous breaking of \ntwoo supersymmetry.

The FI $D$-term (to be denoted $\xi_3$) is not the only way to stabilize the BPS-saturated flux-tube solitons.  In \ntwo
supersymmetric theories one could alternatively introduce it through $F$ terms of the form $\cw=\xi\ca$.
In fact, the FI $D$- and $F$-terms form a triplet of the  global SU(2)$_R$ \cite{HSZ,VY}.
 This explains our notation $\xi_3$ for the
coefficient in front of the FI $D$-term. The FI $F$-term  coefficient $\xi$  is
complex and can be
written as $\xi=\xi_1+i\xi_2$, where $\xi_i$ ($i=1,2,3$) form an SU(2)$_R$ triplet.

In the past we considered the $F$-term stabilized flux tubes e.g. in \cite{proto}. When \ntwo deformations are
introduced in the bulk,\footnote{We mean such deformations that break \ntwo down to \none.} the SU(2)$_R$ symmetry is broken, and the equivalence
between the $D$-term and $F$-term stabilized flux tubes disappears. In particular, the \ntwot-breaking deformation on the world sheet of the $F$-term stabilized strings
does not coincide with that determined in \cite{Edalati,SYhet}. The following question arises:
 Given an \ntwo bulk theory with no Fayet--Iliopoulos term, deformed by an \ntwo
 breaking  superpotential, how can one calculate the corresponding \ntwot-breaking potential
on the $F$-term stabilized string world sheet? 
In the present paper we address this question in the limit of weak deformations
(i.e. in the leading order in the deformation parameters).
We consider U$(N)$ gauge theories with $N_f$
matter hypermultiplets where
we require
\beq
N\leq N_f<2 N\,.
\label{order}
\eeq
The string-stabilizing/deformation terms are introduced via superpotentials, i.e. as $F$-terms.
We find, say, for the U(2) theory with $N_f=3$, that if the bulk deformation is introduced
as
\beq
\cw_{3+1} = \frac{1}{2}\left[\mu_1\,\ca^2 +\mu_2\,\left(\ca^a\right)^2
\right]
\eeq
(i.e. \`a la Seiberg--Witten \cite{SW1,SW2}), the \ntwot-breaking potential it generates
on the string world sheet is
\beq
V_{1+1}(\sigma)=4\pi\left| \mu_1\,m -
\mu_2\left(\sqrt{2}\sigma-\frac{\Lambda}{2}+m\right)\right|,
\label{2Dpotpp}
\eeq
where $m$ is the average (over three flavors) mass term,
\beq
m = \frac{1}{3} \left(m_1 + m_2+m_3\right)\,.
\label{aver}
\eeq
From (\ref{2Dpotpp}) one can read off vacuum energies
for  two vacua of the heterotic weighted CP(2) model at hand.
These vacua correspond to two strings. 
They are BPS-saturated in the effective low-energy U(1) theory. However, if considered in the full theory,
they are non-BPS.\footnote{This is due to the fact that in \none non-Abelian 
gauge theories with $\xi_3=0$ there is no string central charge \cite{gorsh}.
This central charge appears in \ntwo theories provided $\xi_{1,2,3}\neq 0$.} This means that 
\ntwoo supersymmetry is spontaneously broken 
in the weighted CP(1) model we deal with already at the classical level 
(as opposed to the quantum-level  breaking in \cite{Edalati,SYhet}).

To present things in a proper perspective, let us return for a short while to the $D$-term stabilization.
We recall that
the FI $D$-term singles out a particular $r$-vacuum, with $r=N$ (i.e. $N$ quark flavors out of $N_f$
develop 
a vacuum expectation value (VEV)). For instance, one can choose the quark condensate as
\beqn
\bra q^{kA}\ket &=&\sqrt{
\xi_3}\,
\left(
\begin{array}{ccccc}
1 & \ldots & 0,&0&... \\
\ldots & 1 & \ldots,&0& ...\\
0 & \ldots & 1,&0& ...\\
\end{array}
\right),
\nonumber\\[4mm]
k&=&1,..., N\,,\qquad A=1,...,N_f\, ,
\label{qvevD}
\eeqn
where the quark fields are represented in the matrix form, as an $N\times N_f$ matrix in
the color and flavor indices.  Consider the simplest case with
$N_f=N$.
The  vacuum field (\ref{qvevD}) results in  the spontaneous
breaking of both the gauge U($N$) group and flavor (global) SU($N$) group, leaving unbroken a
  diagonal global SU$(N)_{C+F}$,
\beq
{\rm U}(N)_{\rm gauge}\times {\rm SU}(N)_{\rm flavor}
\to {\rm SU}(N)_{C+F}\,.
\label{c+f}
\eeq
Thus, a color-flavor locking takes place in the vacuum.
The presence of the global SU$(N)_{C+F}$ group is a key reason for the
formation of non-Abelian strings whose main feature is
the occurrence  of orientational zero modes associated with rotations of the flux
inside the  SU$(N)_{C+F}$ group. Dynamics of these orientational moduli are described by
the effective two-dimensional   \ntwot supersymmetric CP$(N-1)$ model on
the string world sheet. Next we add the quark mass terms $m_A$ ($A=1,2,..., N$). 
If they are unequal, the global SU$(N)_{C+F}$ group is broken down to U(1)$^{N-1}$
by VEVs of the adjoint fields $\ca^a$. If one assumes that the mass term differences are small, i.e. 
$|m_A-m_B|\ll\sqrt{\xi_3}$,  the orientational moduli, being lifted, remain as quasimoduli.
The  two-dimensional low-energy  theory that emerges in this case on the  world sheet is the
\ntwot supersymmetric CP$(N-1)$ model {\em with twisted masses}. Note, that in this case
the CP$(N-1)$ model still has $N$ degenerate supersymmetric vacua which are identified
with $N$ elementary $Z_N$ strings of the bulk theory, see for example our review \cite{SYrev}.

The \ntwo-breaking  bulk deformations considered in the literature
 \cite{SYnone,Edalati,Tongd,SYhet,SYhetN,BSYhet,BSYhetmass} are as follows:
the  mass term $\mu$ for the adjoint matter
in the theory with non-zero $\xi_3$ and $m_A=0$ (for all $A$) or,  more general superpotentials,
 with the critical points
coinciding with the quark masses $m_A$. 
The reason is rather obvious.
Consider, say, the mass term $\mu$ for the adjoint matter.  If  $m_A=0$, no  FI $F$-terms are generated and the
bosonic parts of the classical string solutions do not depend on $\mu$
\cite{SYnone,SYhet}.  The world sheet-theory changes
only in the fermion sector.
It becomes \ntwoo supersymmetric (heterotic) CP$(N-1)$ model \cite{Edalati,SYhet,BSYhet}.
In the gauged formulation 
 the  deformation potential on the moduli space has the form
\beq
 V_{1+1}=4|\omega|^2\,|\sigma|^2\,,
\label{zeromasspot}
\eeq
where $\sigma$ is a scalar superpartner of the (auxiliary) U(1) gauge field \cite{W93}, while $\omega $
is a world-sheet
deformation parameter determined by the mass $\mu$ of the adjoint fields in the bulk theory,
\beq
\omega \sim \frac{\mu}{\sqrt{\xi_3}}\,,
\label{omega}
\eeq
at small $\mu$ (to the leading order in $\mu$).
In other words, up to an overall normalization, $\cw_{1+1}(\sigma)$ has the same functional form as
$\cw_{3+1}(\ca)$.
 A similar situation takes place at $m_A\neq 0$ provided that the critical points
 of $\cw_{3+1}(\ca)$ are at $m_A$.

In this paper we consider $$\xi_3=0$$ (if not stated to the contrary, in some occasional passages),
both stabilization and \ntwo breaking  are provided by $F$-terms,
 induced by non-zero $\mu$ times quark masses, which are
  are generically considered to be different.
 Now the $Z_N$ strings (their number is $N$) become split. Supersymmetry on the string world sheet
 is spontaneously broken already at the classical level.
We calculate the string tensions in the limit
\beq
\left|\mu/m_A\right|\ll 1\,.
\label{smallmu}
\eeq
 In this limit each of the  $Z_N$ strings is still BPS-saturated in the
associated U(1) low-energy gauge theory arising from the gauge symmetry breaking
U$(N)\to$U(1)$^{N}$ 
through Higgsing.
Of course, all $N$ strings are non-BPS in the full
U$(N)$ gauge theory. 

We find the potential $V_{1+1}(\sigma)$ induced 
in the world-sheet heterotic CP model due to the bulk $\mu$-deformation $\cw_{3+1}(\ca)$.
This potential explicitly exhibits
the breaking of \ntwoo supersymmetry at the classical level and splitting of the energies
of $N$ vacua (the tensions of the $Z_N$ strings).

In our previous works we revealed a number of ``protected" quantities, such as the masses
of the (confined) monopoles. These
parameters are calculable both,
in the bulk theory and on the world sheet, with one and the same result.
The first example of this remarkable correspondence was the explanation \cite{SYmon,HT2} of the
coincidence of the BPS monopole spectrum  in four-dimensional \ntwo supersymmetric QCD in
the $r=N$ vacuum
on the Coulomb branch
at $\xi_3=0$ (given by the exact Seiberg-Witten solution \cite{SW2}), on the one hand,  with the
BPS kink spectrum  in the \ntwot supersymmetric CP
 model, on the other hand. This coincidence was noted in \cite{Dorey,DoHoTo,DTo}.
 The above-mentioned explanation
 \cite{SYmon,HT2} is : (i) the confined monopoles of the bulk theory (represented
 by two-string junctions) are seen as kinks interpolating between two different vacua in the
sigma model on the string world sheet;  (ii) the masses of the BPS monopoles cannot depend on
the nonholomorphic parameter $\xi_3$.

In this paper we find and analyze another example of such exact correspondence between 
the bulk and
world-sheet theories, namely the  tensions of non-Abelian strings stabilized by $F$-terms.
We  study quantum nonperturbative corrections to the string tensions in the bulk theory and show
that they are exactly reproduced by the quantum corrections to vacuum energies in the
heterotic CP model model   on the string world sheet.

The paper is organized as follows.
In Sect.~\ref{bulk} we formulate our theoretical setting -- the bulk \ntwo SQCD with a certain superpotential
which (a) stabilizes the string solutions; (b) breaks \ntwo down to \none. Section~\ref{stringtensions}
is devoted to calculations of the $Z_N$ string tensions in the above bulk theory in the classical limit.
Section~\ref{quantum} deals with (nonperturbative) quantum corrections to the string tensions.
In Sect.~\ref{secWCP} we briefly outline construction of the world-sheet theory in the
limit of unbroken \ntwo in the bulk. In Sect.~\ref{mudeform} we switch on an \ntwo breaking deformation,
and consider its impact on the string world sheet. In Sect.~\ref{2Dquant} (nonperturbative) quantum effects
in the world-sheet theory obtained in Sect.~\ref{mudeform} are analyzed. In Sect.~\ref{monopole}
we revisit the issue of the monopole confinement, i.e. confinement along the string, in addition
to permanent attachment to the strings. This phenomenon is similar to that discussed in
\cite{GSY05}. Section~\ref{genericsup} is devoted to generic single-trace deformation superpotentials 
$\cw_{3+1}(\ca)$. Section~\ref{concl} summarizes our results. In Appendices A and B we discuss details pertinent to particular examples, namely, the U(2) bulk theory with 2 flavors and U(3) theory with 5 flavors, respectively.

\section{Bulk theory}
\label{bulk}
\setcounter{equation}{0}

We start with the description of the bulk theory with which we will
deal throughout the paper.
The gauge symmetry of the basic bulk theory is
U$(N)=$SU$(N)\times$U(1), the number of the matter hypermultiplets in the
fundamental representation is $N_f$. With the deformation superpotential
switched off  this theory has \ntwo supersymmetry. In addition to $N_f$  quark hypermultiplets
(with the mass terms $m_A$, $A=1,2, ..., N_f$) the theory has 
  gauge bosons, gauginos and their  superpartners.
 We assume
$N_f\geq N$ but $N_f<2N$. The latter inequality ensures  asymptotic freedom of the  theory.
Then we will introduce the deformation superpotential of the type $\mu\ca^2$ 
for the adjoint matter breaking \ntwo supersymmetry down to \none. (Later we will consider some other
deformation superpotentials too.)

In more detail, the field content is as follows. The \ntwo vector multiplet
consists of the  U(1)
gauge field $A_{\mu}$ and the SU$(N)$  gauge field $A^a_{\mu}$,
where $a=1,..., N^2-1$, plus their Weyl fermion superpartners
 plus
complex scalar fields $a$, and $a^a$ and their Weyl superpartners.
The $N_f$ quark hypermultiplets of  the U$(N)$ theory consist
of   the complex scalar fields
$q^{kA}$ and $\tilde{q}_{Ak}$ (squarks) and
their   fermion superpartners, all in the fundamental representation of
the SU$(N)$ gauge group.
Here $k=1,..., N$ is the color index
while $A$ is the flavor index, $A=1,..., N_f$. We will treat $q^{kA}$ and $\tilde{q}_{Ak}$
as rectangular matrices with $N$ rows and $N_f$ columns.

As was mentioned, the undeformed theory has
 \ntwo$\!.$ The  superpotential has the form
 \beq
{\mathcal W}_{{\mathcal N}=2} = \sqrt{2}\,\sum_{A=1}^{N_f}
\left( \frac{1}{ 2}\,\tilde q_A {\mathcal A}
q^A +  \tilde q_A {\mathcal A}^a\,T^a  q^A\right)\,,
\label{superpot}
\eeq
where ${\mathcal A}$ and ${\mathcal A}^a$ are  the adjoint chiral superfields, the ${\mathcal N}=2$
superpartners of the gauge bosons of the U(1) and SU($N$) parts, respectively.

Next, we add the mass term for the adjoint fields which, generally speaking,  breaks
supersymmetry down to \none,
\beq
{\mathcal W}_{3+1}=\sqrt{\frac{N}{2}}\,\frac{\mu_1}{2} {\mathcal A}^2
+  \frac{\mu_2}{2}({\mathcal A}^a)^2,
\label{msuperpotbr}
\eeq
where $\mu_1$ and $\mu_2$ is are some mass parameters for the adjoint chiral
superfields, 
U(1) and SU($N$), respectively. The subscript 3+1 tells us that the deformation
superpotential (\ref{msuperpotbr}) refers to the bulk four-dimensional theory.
Clearly, the mass term (\ref{msuperpotbr}) splits \ntwo supermultiplets.

The bosonic part of our basic
theory has the form  (for details see e.g. the review paper \cite{SYrev})
\beqn
S&=&\int d^4x \left[\frac1{4g^2_2}
\left(F^{a}_{\mu\nu}\right)^2 +
\frac1{4g^2_1}\left(F_{\mu\nu}\right)^2
+
\frac1{g^2_2}\left|D_{\mu}a^a\right|^2 +\frac1{g^2_1}
\left|\partial_{\mu}a\right|^2 \right.
\nonumber\\[4mm]
&+&\left. \left|\nabla_{\mu}
q^{A}\right|^2 + \left|\nabla_{\mu} \bar{\tilde{q}}^{A}\right|^2
+V(q^A,\tilde{q}_A,a^a,a)\right]\,.
\label{model}
\eeqn
Here $D_{\mu}$ is the covariant derivative in the adjoint representation
of  SU$(N)$, while
\beq
\nabla_\mu=\partial_\mu -\frac{i}{2}\; A_{\mu}
-i A^{a}_{\mu}\, T^a\,.
\label{defnabla}
\eeq
We suppress the color  SU($N$)  indices of the matter fields. The normalization of the
 SU($N$) generators  $T^a$ is as follows
$$
{\rm Tr}\, (T^a T^b)=\mbox{$\frac{1}{2}$}\, \delta^{ab}\,.
$$
The coupling constants $g_1$ and $g_2$
correspond to the U(1)  and  SU$(N)$  sectors, respectively.
With our conventions, the U(1) charges of the fundamental matter fields
are $\pm1/2$, see Eq.~(\ref{defnabla}).
The scalar potential $V(q^A,\tilde{q}_A,a^a,a)$ in the action (\ref{model})
is the  sum of the $D$ and  $F$  terms,
\beqn
V(q^A,\tilde{q}_A,a^a,a) &=&
 \frac{g^2_2}{2}
\left( \frac{1}{g^2_2}\,  f^{abc} \bar a^b a^c
 +
 \bar{q}_A\,T^a q^A -
\tilde{q}_A T^a\,\bar{\tilde{q}}^A\right)^2
\nonumber\\[3mm]
&+& \frac{g^2_1}{8}
\left(\bar{q}_A q^A - \tilde{q}_A \bar{\tilde{q}}^A -N\xi_3\right)^2
\nonumber\\[3mm]
&+& 2g^2_2\left| \tilde{q}_A T^a q^A
+\frac{1}{\sqrt{2}}\,\,\frac{\pt{\mathcal W}_{3+1}}{\pt a^a}\right|^2+
\frac{g^2_1}{2}\left| \tilde{q}_A q^A +\sqrt{2}\,\,\frac{\pt{\mathcal W}_{3+1}}{\pt a} \right|^2
\nonumber\\[3mm]
&+&\frac12\sum_{A=1}^{N_f} \left\{ \left|(a+\sqrt{2}m_A +2T^a a^a)q^A
\right|^2\right.
\nonumber\\[3mm]
&+&\left.
\left|(a+\sqrt{2}m_A +2T^a a^a)\bar{\tilde{q}}^A
\right|^2 \right\}\,.
\label{pot}
\eeqn
Here  $f^{abc}$ denote the structure constants of the SU$(N)$ group,
$m_A$ is the mass term for the $A$-th flavor,
 and
the sum over the repeated flavor indices $A$ is implied. 
For completeness
we indicated here
the FI $D$-term  $\xi_3$. As was mentioned, 
in the bulk of the paper $\xi_3$ is set at zero. Only occasionally we 
make digressions and discuss  $\xi_3\neq 0$. In these cases it is clearly stated that $\xi_3\neq 0$.

The vacuum structure of  this theory is as follows.
The  vacua of the theory (\ref{model}) are determined by zeros of
the potential (\ref{pot}). In the general case, the theory has many so-called $r$-vacua,
i.e. those vacua in which
 $r$ quarks condense, where $r$ can take any value up to $N$, $r=0,...,N$. Say, $N$  vacua with $r=0$
 are always at strong
coupling. These are the   monopole   vacua of Ref.~\cite{SW1,SW2}. 

We   focus
on a particular set of vacua with the maximal number of condensed quarks, i.e. $r=N$.
The reason for this choice is that all U(1) factors of the gauge group are spontaneously broken in these vacua, and
they support non-Abelian strings \cite{HT1,ABEKY,SYmon,HT2}.

Let us assume first that our theory is at   weak coupling and thus can
be analyzed quasiclassically. Below we will explicitly formulate necessary  conditions for the quark mass terms
and $\mu$ which will guarantee this regime.
For generic values of the quark masses we have
$$C_{N_f}^{N}= \frac{N_f!}{N!(N_f-N)!}$$ isolated $r$-vacua where 
in our case $r=N$; i.e. $N$  quarks (out of $N_f$) develop
vacuum expectation values
(VEVs).
Consider, say, the (1,2,...,$N$) vacuum in which the first $N$ flavors develop VEVs.
In this vacuum  the
adjoint fields  develop
VEVs too, namely,
\beq
\left\langle \left(\frac12\, a + T^a\, a^a\right)\right\rangle = - \frac1{\sqrt{2}}
 \left(
\begin{array}{ccc}
m_1 & \ldots & 0 \\
\ldots & \ldots & \ldots\\
0 & \ldots & m_N\\
\end{array}
\right),
\label{avev}
\eeq
For generic values of the quark mass parameters, the  SU$(N)$ subgroup of the gauge
group is
broken down to U(1)$^{N-1}$. However, in the special limit
\beq
m_1=m_2=...=m_{N_f},
\label{equalmasses}
\eeq
the  SU$(N)\times$U(1) gauge group remains  unbroken by the adjoint field.
In this limit the theory acquires also a global flavor SU$(N_f)$ symmetry.

With all $m_A$'s equal and to the {\em leading order} in $\mu$ the mass term for the adjoint matter (\ref{msuperpotbr})
reduces to the FI $F$-term of the U(1) factor $\xi \sim \mu m/\sqrt{N}\,$ (see below)
  which does not break \ntwo supersymmetry \cite{HSZ,VY}. In this case the FI $F$-term
 can be transformed into the FI $D$-term by an SU$(2)_R$ rotation, and the theory reduces to
\ntwo  supersymmetric QCD described e.g. in the review \cite{SYrev}.
Higher orders in parameter
(\ref{smallmu}) break \ntwo supersymmetry explicitly splitting \ntwo supermultiplets.

If the values of $m_A$ are unequal, the   U($N$) gauge group is spontaneously 
broken  down to U(1)$^{N}$
by VEVs of $a^a$, see  (\ref{avev}). To the leading order in $\mu$ the superpotential
in (\ref{msuperpotbr}) reduces to $N$ distinct FI terms: one $F$-term for 
 each U(1) gauge factor. Thus,
\ntwo supersymmetry in each individual low-energy U(1) sector is unbroken. It gets
broken, however, being considered in the full microscopic U($N$) gauge theory.

Assuming that $\xi_3=0$ and using (\ref{msuperpotbr}) and (\ref{avev}) we get from 
(\ref{pot}) all VEVs of the squark fields. Upon gauge rotation they can be written in the form
\beqn
\langle q^{kA}\rangle &=& \langle\bar{\tilde{q}}^{kA}\rangle=\frac1{\sqrt{2}}\,
\left(
\begin{array}{cccccc}
\sqrt{\xi_1} & \ldots & 0 & 0 & \ldots & 0\\
\ldots & \ldots & \ldots  & \ldots & \ldots & \ldots\\
0 & \ldots & \sqrt{\xi_N} & 0 & \ldots & 0\\
\end{array}
\right),
\nonumber\\[4mm]
k&=&1,..., N\,,\qquad A=1,...,N_f\, ,
\label{qvev}
\eeqn
where  the quark fields are represented by matrices carrying color and flavor indices.
Here we define the FI $F$-term parameters for each U(1) gauge factor as follows
\beq
\xi_P=2\left\{\sqrt{\frac{2}{N}}\,\mu_1\,m+\mu_2(m_P-m)\right\},
\qquad P=1, ..., N\,.
\label{xis}
\eeq
Moreover,  $m$ is the average value of the first $N$ quark masses,
\beq
m=\frac1{N}\sum_{P=1}^{N} m_P.
\label{avm}
\eeq

While the adjoint field condensation does not break the SU$(N)\times$U(1) gauge group in the limit
(\ref{equalmasses}), in the very same limit the quark condensate (\ref{qvev}) results in the spontaneous
breaking of both gauge and flavor symmetries.
A diagonal global SU$(N)$ combining the gauge SU$(N)$ and an
SU$(N)$ subgroup of the flavor SU$(N_f)$
group survives, however. Below we will refer to this diagonal
global symmetry as to color-flavor locked $ {\rm SU}(N)_{C+F}$.

More exactly, the pattern of the spontaneous breaking of the
color and flavor symmetry
is as follows:
\beq
{\rm U}(N)_{\rm gauge}\times {\rm SU}(N_f)_{\rm flavor}\to  {\rm SU}(N)_{C+F}\times  {\rm SU}(\tilde{N})_F\times {\rm U}(1)\,,
\label{c+fxf}
\eeq
where $\tilde{N}=N_f-N$.
Here  the SU$(\tN )_F$ factor represents the flavor rotation of the ``extra"
$\tN$ quarks.
The phenomenon of color-flavor locking in the case at hand is slightly different  than 
that in the case $N_f =N$.
The presence of the global SU$(N)_{C+F}$ group is instrumental for
formation of the non-Abelian strings (see below).

For unequal quark mass parameters both the adjoint and squark VEVs 
brake  the  global symmetry  (\ref{c+fxf}) down
  to
U(1)$^{N_f-1}$. This should be contrasted with the
theory with the FI $D$-term ($\xi_3\neq 0$, $\mu_1=\mu_2=0$) in which
the squark VEVs are all equal and do not break color-flavor symmetry.

The above quasiclassical analysis is valid if the theory is at weak coupling.
This is the case  if the
mass differences are large,
\beq
|\Delta m_{AB}|\equiv |m_A-m_B|\gg \Lambda\,,
\label{U1N}
\eeq
 or the quark VEV are large
while the mass differences can be not-so-large. In the first case
the theory at low energies reduces to U(1)$^N$ gauge theory. In the second case it remains
U$(N)$ gauge theory in which the coupling constant is frozen at the scale equal to
the large values of the
quark condensate.

From (\ref{qvev}) we see that the quark condensates are of 
the order of
$\sqrt{\mu m}$, see also \cite{SW1,SW2,APS,CKM} 
(we assume that $\mu_1\sim\mu_2\sim\mu$). In this case the weak
coupling condition is
\beq
\sqrt{\mu m}\gg\Lambda\, .
\label{weakcoup}
\eeq
We assume that at least one of  conditions (\ref{U1N}) or (\ref{weakcoup}) is fulfilled.
In particular,  the condition (\ref{weakcoup}) combined with the condition of small $\mu$ (\ref{smallmu})  ensures that the average quark mass $m$
is very large. In the theory with the
FI $D$-term the average quark mass can  always assumed to
vanish by virtue of a  shift of the adjoint U(1) field. In the case 
under consideration, when $\xi_3=0$ and stabilization is achieved through $F$-terms,
the presence of the  deformation
(\ref{msuperpotbr}) forbids this shift; hence,  $m$ becomes a physical parameter.

In fact, we can relax both  conditions (\ref{U1N}) and  (\ref{weakcoup}) and
pass to the strong coupling domain at
\beq
\sqrt{\mu m}\ll\Lambda, \qquad |\Delta m_{AB}|\ll \Lambda\,
\label{strcoup}
\eeq
by virtue of duality.
We demonstrated  \cite{SYdual,SYtorkink,SYcross} that in this passage the theory goes through
crossover transitions; in the  domain (\ref{strcoup}) it
can be described in terms of a weakly coupled (non-asymptotically free) dual theory
with with the gauge group
\beq
{\rm U}(\tN)\times {\rm U}(1)^{N-\tN}\,,
\label{dualgaugegroup}
\eeq
 and $N_f$ flavors of light {\em dyons}.\footnote{This non-Abelian  duality is similar to Seiberg's
 duality in \none supersymmetric QCD
\cite{Sdual,IS}. Also a dual non-Abelian gauge group SU$(\tN)$ was identified on the Coulomb branch
at the root of a baryonic Higgs branch in the \ntwo supersymmetric  SU($N$) gauge theory with massless quarks \cite{APS}.} We will see that our results for non-Abelian string tensions,
as well as the effective world-sheet theory obtained at weak coupling, 
can be analytically continued into the domain (\ref{strcoup}).

\section{\boldmath{$Z_N$} string tensions}
\label{stringtensions}
\setcounter{equation}{0}

In Sect.~\ref{bulk} we argued that the quark fields develop VEVs in the $r=N$ vacuum which break 
the gauge group, see (\ref{qvev}). Therefore, our theory  supports strings.
In fact, the minimal stings in our theory are the $Z_N$ strings, progenitors of 
the non-Abelian strings,
having the U(1) field fluxes reduced by the factor $1/N$ compared to
that of the Abrikosov--Nielsen--Olesen
\cite{ANO} string. In these $Z_N$ strings the squark fields have windings both in the U(1) and SU$(N)$
gauge factors \cite{HT1,ABEKY,SYmon,HT2}.

We will study these strings applying the same methods as those used in the \ntwo theory with
the FI $D$-term, or with one common FI $F$-term (which can be transformed into the $D$-term by
virtue of an SU$(2)_R$
rotation),  see the review \cite{SYrev}. Here we will be interested only in the tensions of the $Z_N$
strings rather than in full solutions. Therefore, we need to know only the behavior of the gauge and squark fields
at infinity. Using the {\em ansatz}
\beq
 q^{kA} = \bar{\tilde{q}}^{kA}=\frac{\vp^{kA}}{\sqrt{2}}\,,
\label{qtq}
\eeq
and setting the adjoint scalars at their vacuum values  (\ref{avev}) (see \cite{SYrev})
we determine the behavior at $r\to\infty$  of the first of the $Z_N$ strings. We have
\beqn
 \vp^{kA}\ &=& \frac{1}{\sqrt{2}}\,
\left(
\begin{array}{cccccc}
\sqrt{\xi_1}\,e^{i\alpha} & \ldots & 0 & 0 & \ldots & 0\\
0 & \sqrt{\xi_2} & \ldots & 0 &  \ldots & 0\\
\ldots & \ldots & \ldots  & \ldots & \ldots & \ldots\\
0 & \ldots & \sqrt{\xi_N} & 0 & \ldots & 0\\
\end{array}
\right),
\label{qinf}
\eeqn
for the squark fields and
\beqn
 \left(\frac12\, A_i + T^a\, A_{i}^a\right)&=&
\left(
\begin{array}{cccccc}
1 & 0 &\ldots &  0 \\
0 & 0 & \ldots & 0 \\
\ldots & \ldots & \ldots \\
0 & 0 & \ldots  & 0\\
\end{array}
\right)\,\pt_i \alpha,
\label{ginf}
\eeqn
for the gauge fields,
where $r$ and $\alpha$ are the polar coordinates in the (1,\,2) plane orthogonal to the string
axis, $i=1,\,2$. Asymptotic behavior of other $Z_N$ strings is obtained by
assigning the winding factor $e^{i\alpha}$ to any other diagonal  element in the  matrix
(\ref{qinf}) and putting the corresponding diagonal element of (\ref{ginf})  to
unity.
Equation~(\ref{ginf}) implies that the flux of the $P$-th   string is
\beqn
 \left(\frac12\, F_3^{*} + T^a\, F_3^{*a}\right)&=&
2\pi\left(
\begin{array}{cccccc}
0  &\ldots &  0 \\
\ldots & \ldots & \ldots \\
  \ldots & 1 & \ldots  \\
\ldots & \ldots & \ldots \\
0 &  \ldots  & 0\\
\end{array}
\right)\, ,
\label{flux}
\eeqn
where the only nonvanishing element (equal to unity) is located at the diagonal of the matrix
above at the $P$-th row.

Assuming all $m_A$'s to be different and putting  all off-diagonal 
squark and gauge fields to zero we  reduce our non-Abelian theory (\ref{model})
to the Abelian U(1)$^N$ gauge theory.
Then, assuming   profile functions of the string solutions to be dependent only
on $x_i$ ($i=1,\,2$), we can write the Bogomol'nyi representation \cite{B} for
the $Z_N$ string tensions. From (\ref{model}) we have (cf. \cite{SYrev})
\beqn
T
&=&
\int{d}^2 x   \left\{
\left[\frac1{\sqrt{2}g_2}F^{*a}_{3} +
\frac{g_2}{\sqrt{2}}
\left(\bar{\vp}_A T^a \vp^A + \sqrt{2}\mu_2\langle a^a\rangle\right)\right]^2
\right.
\nonumber\\[3mm]
&+&
\left[\frac1{\sqrt{2}g_1}F^{*}_{3} +
\frac{g_1 }{2\sqrt{2}}
\left(|\vp^A|^2+ 2\sqrt{N}\mu_1\langle a\rangle\right)\right]^2
\nonumber\\[5mm]
&+&
\left.
 \left|\nabla_1 \,\vp^A +
i\nabla_2\, \vp^A\right|^2
 -\sqrt{N}\mu_1\langle a\rangle\,  F^{*}_3
-\sqrt{2}\mu_2\langle a^a\rangle\,  F^{*a}_3\right\},
\label{bogs}
\eeqn
where $$F_{3}^{*}=\, F_{12}\,\,\,
\mbox{and}\,\,\, F_{3}^{*a}=\, F_{12}^{a}\,.$$
In the above expression $\langle a\rangle$ and $\langle a^a\rangle$ are just numbers given
by (\ref{avev}). This is the low-energy approximation which reduces the superpotential (\ref{msuperpotbr})
to individual FI terms in each of the U(1) gauge. In this approximation the $Z_N$ strings are BPS saturated.

The Bogomol'nyi representation (\ref{bogs})
leads us to the following first-order equations:
\beqn
&& F^{*}_{3}+\frac{g_1^2}{2} \left(\left|
\varphi^{A}\right|^2 + 2\sqrt{N}\mu_1\langle a\rangle\right)\,=0\, ,
\nonumber\\[3mm]
&& F^{*a}_{3}+ g_2^2 \left(\bar{\vp}_{A}T^a \varphi^{A}+ \sqrt{2}\mu_2\langle a^a\rangle\right)
\,=0\, ,
\nonumber\\[3mm]
&& (\nabla_1+i\nabla_2)\varphi^A=0\, .
\label{gfoes}
\eeqn
Once these equations are satisfied, the energy of the BPS object is
given by  two last surface terms in (\ref{bogs}).\footnote{Note that the representation
(\ref{bogs}) can be written also with the opposite  sign in front of
the flux terms. Then we would get the Bogomol'nyi equations for  antistring.}
Substituting (\ref{avev}) and the gauge field fluxes  (\ref{flux})
for each of the $Z_N$ string in the two last surface terms  in (\ref{bogs}) we arrive at the
following tensions:
\beq
T^{\rm BPS}_{P}=2\pi|\xi_P|, \qquad P=1,...,N,
\label{ten}
\eeq
where $T^{\rm BPS}_{P}$ is the tension of the string associated with the winding
of the $P$-th quark (see (\ref{qinf})), while
$N$ complex FI $F$-terms $\xi_P$ are classically determined by $\mu$'s and $m$'s via Eq.~(\ref{xis}).
We see that the tension of the $P$-th elementary string is determined
by the  condensate of the very same squark that  winds at infinity (cf. (\ref{qvev})).

As longs as the string solitons are BPS saturated, their tensions must be given by exact expressions.
Equation (\ref{xis}) gives the FI parameters in the semiclassical
approximation.
Later we will see that there are nonperturbative corrections to $\xi_P$'s that are   
 $O(\Lambda/m_A)$. The $\mu$ dependence is a different story,
however. The tensions of the BPS saturated $Z_N$ strings
 are presented in (\ref{ten}) only to the leading (linear) order in $\mu$.
Higher orders in $\mu$ destroy the property
that the superpotential (\ref{msuperpotbr}) is representable as an FI term;  they also  break \ntwo supersymmetry       
making strings non-BPS saturated.


\section{Quantum effects}
\label{quantum}
\setcounter{equation}{0}

This section is devoted to  calculation of  (nonperturbative) quantum  corrections
 $O(\Lambda/m_A)$ to the $Z_N$ string tensions. The idea is straightforward: 
 we exploit the exact Seiberg--Witten solution of the theory on the Coulomb branch
\cite{SW1,SW2} (more exactly, the SU$(N)$ generalizations
of the Seiberg--Witten solution \cite{ArFa,KLTY,ArPlSh,HaOz})
to calculate the FI $F$-terms $\pt{\mathcal W}_{3+1}/\pt a^a$ and
$\pt{\mathcal W}_{3+1}/\pt a$ {\em exactly}  rather than in the semiclassical approximation,
as in Sects.~\ref{bulk} and \ref{stringtensions}.
Defining
\beq
u_k= \bra {\rm Tr}\left(\frac12\, a + T^a\, a^a\right)^k\ket, \qquad k=1, ..., N\,,
\label{u}
\eeq
we perform a quantum generalization in  the two relevant terms in the third line in (\ref{pot}),
\beq
\frac{\pt{\mathcal W}_{3+1}}{\pt a^a}\to\mu_2 \,\frac{\pt u_2}{\pt a^a}\,,
\qquad \quad\frac{\pt{\mathcal W}_{3+1}}{\pt a}\to\mu_1 \sqrt{\frac{2}{N}}\,\frac{\pt u_2}{\pt a}\,.
\label{dwda}
\eeq
Then the two last surface terms in (\ref{bogs}) take the form
\beq
T ^{\rm BPS}=-\int{d}^2 x
\left\{\mu_1 \frac{2}{\sqrt{N}}\,\frac{\pt u_2}{\pt a}  F^{*}_3
+\sqrt{2}\mu_2\,\frac{\pt u_2}{\pt a^a}\,  F^{*a}_3\right\},
\label{quantfluxterms}
\eeq
where $\pt u_2/\pt a$ and $\pt u_2/\pt a^a$ in the $r=N$ vacuum 
under consideration are some constants depending on $m_A$ and $\Lambda$.

Clearly only the diagonal gauge fluxes do not vanish.
Substituting the gauge fluxes from (\ref{flux}) in (\ref{quantfluxterms}) we get
\beq
T ^{\rm BPS}_P=4\pi\sqrt{2}\,\left|\sqrt{\frac{2}{N}}\,\mu_1\,e+\mu_2(e_P-e)\right|,
\label{qten}
\eeq
where $e_P$ ($P=1, ..., N$) are the diagonal elements of the $N\times N$ matrix
\beq
E=\frac1{N}\,\frac{\pt u_2}{\pt a}+T^{\tilde{a}}\,\frac{\pt u_2}{\pt a^{\tilde{a}}}\,,
\label{E}
\eeq
$T^{\tilde{a}}$ are the Cartan generators of the SU$(N)$ gauge group ($\tilde{a}=1,...,(N-1)$),
 while $e$ is their average value,
\beq
e=\frac1N\sum_{P=1}^{N} e_P\,.
\label{e}
\eeq
If $\Delta m_{AB}\gg \Lambda$, we can rely on the quasiclassical expressions namely, the matrix $E$ reduces to the matrix
 (\ref{avev}) and 
\beq
\sqrt{2}e_P\approx-m_P, \qquad \sqrt{2}e\approx-m\,,
\label{classe}
\eeq
and we then immediately recover the classical result (\ref{ten}) from (\ref{qten}) where $\xi_P$'s are given in Eq.~(\ref{xis}).

Our current task is to find $e_P$'s from the exact solution of the theory on the Coulomb branch
(at $\mu=0$).
The Seiberg--Witten curve in the theory under consideration has the form \cite{APS}
\beq
y^2= \prod_{P=1}^{N} (x-\phi_P)^2 -
4\left(\frac{\Lambda}{\sqrt{2}}\right)^{N-\tN}\, \,\,\prod_{A=1}^{N_f} \left(x+\frac{m_A}{\sqrt{2}}\right),
\label{curve}
\eeq
where $\phi_P$ are gauge invariant parameters on the Coulomb branch. Their relations to the
gauge invariant parameters $u_k$ in (\ref{u}) are as follows:
\beq
u_k=\sum_{P=1}^{N} \phi^k_P\,.
\label{uphi}
\eeq
The   curve (\ref{curve}) describes the Coulomb branch of the theory for  $N_f<2N-1$. The case  $N_f=2N-1$ 
(i.e. $   \tN=N-1$) is special. In this case we must make a shift \cite{APS}
\beq
m_A\to \tilde{m}_A=m_A+\frac{\Lambda}{N}, \qquad N_f=2N-1\,,
\label{shift}
\eeq
in (\ref{curve}).

\vspace{1mm}

Semiclassically,
at large masses
\beq
{\rm diag}\left(\frac12\, a + T^a\, a^a\right) \approx
\left[\phi_1,...,\phi_N\right]\,.
\eeq
Therefore, in  the ($1, ... ,\, N$) quark vacuum we have
\beq
\phi_P \approx -\frac{m_P}{\sqrt{2}},\qquad P=1, ... ,\, N\,,
\label{classphi}
\eeq
in the large $m_A$ limit, see (\ref{avev}).
In terms of the curve (\ref{curve})
this vacuum corresponds to
such values of $\phi_P$ which ensure that the curve has $N$ double roots and
$\phi_P$'s are determined by the quark mass parameters in the semiclassical limit, see (\ref{classphi}).
The presence of $N$ double roots means that $N$ quark flavors are massless at this singular
point on the Coulomb branch. Upon $\mu$ deformation this singularity becomes the $r=N$
vacuum, where the  $N$ ``former" massless squarks condense.

The curve (\ref{curve}) with $N$ double roots has the  form
\beq
y^2= \prod_{P=1}^{N} (x-e_P)^2,
\label{rNcurve}
\eeq
where semiclassically $e_P$'s are given  by mass parameters via (\ref{classe}).
In Appendices A and B we
show  that the double roots $e_P$ of the Seiberg--Witten curve are precisely given by  the diagonal
elements of the matrix $E$ (\ref{E}). Thus, the Seiberg--Witten curve
 which appeared in \cite{SW1,SW2} as a kind of an auxiliary mathematical tool
acquires a physical and very transparent meaning in the $r=N$ vacuum: its double roots
determine the tensions of the $Z_N$ strings through Eq.~(\ref{qten})!

In Appendices A and B we demonstrate that diagonal elements of matrix $E$ are given by
$N$ double roots of the Seiberg--Witten curve (\ref{curve}) by considering two
simple examples:  $N_f=N$ and $2N>N_f>N$ theories. More exactly, we
analyze there the $N_f=N=2$ and $N=3$, $N_f=5$
cases. Say, for the simplest $N_f=N=2$ case, the two double roots of (\ref{curve}) are
as follows \cite{SYcross}:
\beq
\sqrt{2}\, e_{1,2}=-m \mp \sqrt{\frac{\Delta m_{12}^2}{4} +\Lambda^2}\, .
\label{eN2}
\eeq
Note that the average value of $e$ is proportional to the average mass $m$,
\beq
\sqrt{2}e=-m\, ,
\label{em}
\eeq
where $e$ and $m$ are given in (\ref{e}) and (\ref{avm}),
 respectively. In fact, this property is fulfilled for all cases with $N_f\le 2N-1$.

Substituting (\ref{eN2}) into (\ref{qten}), we obtain the tensions of two $Z_2$ strings,
\beq
T ^{\rm BPS}_{1,2}=4\pi\,\left|\mu_1\,m \pm \mu_2 \sqrt{\frac{\Delta m_{12}^2}{4}
 +\Lambda^2}\,\right|.
\label{qten20}
\eeq
This formula takes into account all quantum (instanton) corrections in powers of $\Lambda/\Delta m_{12}$.

As was already mentioned, one can relax the weak-coupling conditions (\ref{U1N}) and
(\ref{weakcoup}) and enter the strong-coupling domain (\ref{strcoup}). The theory undergoes
a crossover transition, i.e. passes the curves of marginal stability (CMS), where certain states decay,
as well as  monodromies,
which change charges of other states. In this domain the original theory
 can be better described by the dual weakly coupled  theory
of light dyons with the gauge group (\ref{dualgaugegroup}),
derived in \cite{SYcross}.

In Appendices A and B we prove that our result for the $Z_N$-string tensions   (\ref{qten}) can be analytically
continued (as functions of $\Delta m_{AB}$ and $m$) into the strong-coupling domain.
This was expected, of course. We encounter a crossover at $\Delta m_{AB}\sim \Lambda$ and
$\mu m\sim \Lambda^2$ rather than a phase transition \cite{SYcross,SYdual,SYtorkink}. Therefore, string
tensions exhibit a continuous behavior across the crossover lines.

\section{\boldmath{\ntwot\!\!}-supersymmetric world sheet theory}
\label{secWCP}
\setcounter{equation}{0}

In this part of the paper we proceed from the analysis in the bulk to the analysis on the string
world sheet, with the intension to demonstrate that 
both lead to identical consequences. As previously, to establish the appropriate setting,
we start from the undeformed case.
We   briefly review the world-sheet low-energy sigma models on the non-Abelian strings in \ntwo supersymmetric QCD 
with the FI term \cite{HT1,ABEKY,SYmon,HT2}, see also the review papers
\cite{Trev,Jrev,SYrev,Trev2}. First we will deal with \ntwo QCD with the FI $D$-term. The corresponding Lagrangian
is 
(\ref{model}) with $\xi_3\neq 0$ and  large, and
\beq
\mu_1=\mu_2=0\,.
\label{mu0}
\eeq
To begin with, assume that $N_f=N$.
The Abelian $Z_N$-string solutions break the  SU$(N)_{C+F}$ global group down to
${\rm SU}(N-1)\times {\rm U}(1)$. As a result,
the non-Abelian strings develop orientational zero modes associated with rotations of their color
flux inside the non-Abelian SU($N$) group.
The moduli space of the non-Abelian string is described by the coset space
\beq
\frac{{\rm SU}(N)}{{\rm SU}(N-1)\times {\rm U}(1)}\sim {\rm CP}(N-1)\,,
\label{modulispace}
\eeq
in addition to $C$ spanned by the translational modes. The translational moduli totally decouple.
They are sterile free fields which can be ignored in further considerations.
Therefore, the
low-energy effective theory on the   non-Abelian string
is the two-dimensional \ntwot   CP$(N-1)$ model \cite{HT1,ABEKY,SYmon,HT2}.

Now  let us add ``extra" quark flavors, with degenerate masses, increasing $N_f$ from $N$ 
up to a certain value $N_f >N$ but $N_f\leq 2N-1$.
The strings supported by such theory are {\em semilocal}.
In particular, the string solutions on the Higgs branches (typical
for multiflavor theories) usually are not fixed-radius strings, but, rather,
possess radial moduli, a.k.a size moduli, see    \cite{AchVas} for a comprehensive review of
the Abelian semilocal strings. The transverse size of such a string is not fixed: it can vary without changing the tension.

Non-Abelian semilocal strings in \ntwo SQCD with $N_f>N$ were studied in
\cite{HT1,HT2,SYsem,Jsem}.
The orientational
zero modes of the semilocal non-Abelian string are parametrized by a complex vector $n^P$ ($P=1, ..., N$),
 while its $\tN=(N_f-N)$ size moduli are parametrized by another complex vector
$\rho^K$ ($K=N+1, ..., N_f$). The effective two-dimensional theory
which describes the internal dynamics of the non-Abelian semilocal string is
the \ntwot weighted  CP model on a ``toric" manifold, which includes both types of fields. The bosonic
part of the action
in the gauged formulation (which assumes taking the limit $e^2\to\infty$)
has the form\,\footnote{Equation (\ref{wcp}) and similar expressions below are given in the Euclidean notation.}
\beqn
&&S = \int d^2 x \left\{
 \left|\nabla_{\alpha} n^{P}\right|^2
 +\left|\tilde{\nabla}_{\alpha} \rho^K\right|^2
 +\frac1{4e^2}F^2_{\alpha\beta} + \frac1{e^2}\,
\left|\pt_{\alpha}\sigma\right|^2
\right.
\nonumber\\[3mm]
&+&\left.
2\left|\sigma+\frac{m_P}{\sqrt{2}}\right|^2 \left|n^{P}\right|^2
+ 2\left|\sigma+\frac{m_{K}}{\sqrt{2}}\right|^2\left|\rho^K\right|^2
+ \frac{e^2}{2} \left(|n^{P}|^2-|\rho^K|^2 -2\beta\right)^2
\right\},
\nonumber\\[4mm]
&&
P=1,...,N\,,\qquad K=N+1,...,N_f\,.
\label{wcp}
\eeqn
The fields $n^{P}$ and $\rho^K$ have
charges  +1 and $-1$ with respect to the auxiliary U(1) gauge field;
hence, the corresponding  covariant derivatives in (\ref{wcp}) are defined as
\beq
 \nabla_{\alpha}=\d_{\alpha}-iA_{\alpha}\,,\qquad
\tilde{\nabla}_{\alpha}=\d_{\alpha}+iA_{\alpha}\,,
\label{covarder}
\eeq
  respectively. This is the effective low-energy theory on the non-Abelian string
in the $r=N$ vacuum, in which the first $N$ squark flavors ($P=1, ..., N$) condense.

If only the charge $+1$ fields $n$ were present, in the limit
$e^2\to\infty$ we would get a conventional twisted-mass deformed
CP $(N-1)$ model.
The presence of  the charge $-1$ fields $\rho^K$ converts the CP$(N-1)$
 target space   into that of the a weighted
CP$(N_f-1)$ model.
In parallel to the CP$(N-1)$ model, small mass differences
$\left| m_A-m_B\right|$ lift orientational and size zero modes generating a shallow 
potential on the modular space.
The $D$-term condition
\beq
  |n^P|^2 - |\rho^K|^2=2\beta
\label{unitvec}
\eeq
is implemented in the limit $e^2\to\infty$. Moreover, in this limit
the gauge field $A_{\alpha}$  and its \ntwo bosonic superpartner $\sigma$ become
auxiliary and can be eliminated through equations of motion.

The  two-dimensional coupling constant $\beta$ is related to the four-dimensional
one as (e.g. \cite{SYrev})
\beq
\beta= \frac{2\pi}{g_2^2}\,.
\label{betag}
\eeq
This relation  is obtained  at the classical level
\cite{ABEKY,SYmon}.
 In the quantum theory
both couplings run. In particular, the  model (\ref{wcp}) is asymptotically free
\cite{W93} and develops its own scale, which coincides with that the bulk theory   $\Lambda$
\cite{SYmon}.
The ultraviolet cut-off in the sigma model on the string world sheet
is determined by  $g_2\sqrt{\xi_3}$.
Equation~(\ref{betag}) relating the two- and four-dimensional couplings
is valid at this scale.
At $N\leq N_f<2N$ the  model (\ref{wcp}) is asymptotically free. Its coupling $\beta$
continues running below $g_2\sqrt{\xi_3}$ until it ceases to run and freezes
 at the scale of the mass differences
$|\Delta m_{AB}|$.
 If all mass differences are large,
$|\Delta m_{AB}| \gg \Lambda$, the model is at weak
coupling. From (\ref{wcp}) we see that in this regime the model has $N$ vacua (i.e. $N$
strings from the standpoint of the bulk theory) at
\beq
 \sqrt{2}\sigma=-m_{P_0},\qquad |n^{P_0}|^2=2\beta\,,\qquad n^{P\neq P_0}=\rho^K=0\,,
\label{classvac}
\eeq
where $P_0=1, ..., N$.

\section{Switching on a weak \boldmath{$\mu$}-deformation}
\label{mudeform}
\setcounter{equation}{0}

Let us break \ntwo supersymmetry in the bulk theory by switching on the deformation
superpotential of the type (\ref{msuperpotbr}), assuming that the $\mu$
parameters are small, (\ref{smallmu}). If the parameters 
$m$ are large enough, we can switch off the FI $D$-term parameter $\xi_3$, keeping the theory
in the weak-coupling regime, see (\ref{weakcoup}). 
The string solutions will be stabilized by $F$-terms.
Our aim in this section is to
find an effective low-energy theory on the world sheet of the non-Abelian strings in the deformed case.
If the typical scale of excitations in the world-sheet theory (it is of the order of
max$(\Delta m_{AB},\Lambda)$) is much less than the inverse thickness of the string
$\sim \sqrt{\mu m}$, we can expect that such low-energy world-sheet description exists and
is given by a certain deformation of the \ntwot supersymmetric CP model (\ref{wcp})
which breaks \ntwot down to \ntwoo.
As was already mentioned in Sect.~\ref{intro}, this problem is    solved in the theories with the $D$-term stabilization.
Namely, if we keep $\xi_3$ nonvanishing (and large) and switch on deformation (\ref{msuperpotbr})
putting all   masses   $m_A=0$, then the effective theory on the non-Abelian string
becomes  \ntwoo supersymmetric CP model  with the quadratic in $\sigma$ 
superpotential \cite{Edalati,SYhet,BSYhet}. After a brief review of this result we move on to consideration
of the case we are interested in in this paper: $\xi_3=0$, while $m\neq 0$ and $\Delta m_{AB}\neq 0$.
Switching on  $\mu_{1,2}\neq 0$ generates the FI
$F$-terms in each of the U(1) factors of the
U($N$) gauge group and, simultaneously, breaks \ntwot down to \ntwoo on the world sheet.

\subsection{\boldmath{$|m_A-m_B| = 0$}}
\label{masth}

With four supercharges of the  deformed \none bulk theory normally
the 1/2 BPS-saturated string solution  will preserve
only two supercharges on the string world sheet. However, it is well-known that the sigma model with the CP (K\"{a}hler) target
space, when supersymmetrized, automatically yields \ntwot sigma model; one cannot
get \ntwoo$\!$.
It was
pointed out \cite{Edalati} that the target space in the problem at hand is in fact ${\rm CP}(N_f-1)\times C$
rather than ${\rm CP}(N_f-1)$. Edalati and Tong suggested that
the superorientational zero modes can mix with the supertranslational ones.
They
explicitly constructed an \ntwoo supergeneralization of the sigma model with
the target space ${\rm CP}(N_f-1)\times C$.
In their construction the bosonic part of the \ntwot model (\ref{wcp})
is supplemented by the term
\beq
\delta S_{1+1}= \int d^2 x\,V_{1+1}(\sigma)=\int d^2 x\,
\left|\frac{\pt {\mathcal W}_{1+1}}{\pt \sigma}
\right|^2
\label{etong}
\eeq
breaking \ntwot down to \ntwoo$\!$. Here ${\mathcal W}_{1+1}(\sigma)$ is a two-dimensional deformation superpotential.

Later in \cite{SYhet,BSYhet} this conjecture was confirmed. It was shown that the two-dimensional
superpotential ${\mathcal W}_{1+1}$ is indeed generated on the world sheet of the non-Abelian string in the massless theory.
For the bulk deformation (\ref{msuperpotbr})   the world-sheet superpotential is
\beq
{\mathcal W}_{1+1}=\omega\,\sigma^2,
\label{W1+1massless}
\eeq
 where the deformation parameter is proportional to $\mu$ to the leading order at small  $\mu$.
At  large $\mu$
 the world-sheet deformation
becomes more complicated \cite{SYhet,BSYhet}.
For the superpotential (\ref{W1+1massless}) the scalar potential is   $|\sigma|^2$,
see ({\ref{zeromasspot}).

The massless heterotic \ntwoo supersymmetric \cpn model  with the deformation potential ({\ref{zeromasspot})
was solved in \cite{SYhetN,BSYhetmass} in the large-$N$ approximation. It was shown that, although
classically the model has \ntwoo supersymmetry, it gets spontaneously broken by quantum
non-perturbative effects. On the other hand, the model has $N$ strictly degenerate vacua
with vacuum energies proportional to $\Lambda$.

The vacuum energy in the world-sheet theory is obviously identified with the string tensions
in the bulk theory. Therefore, our result for the string tensions (\ref{ten}) in the $|m_A-m_B|\neq 0$
theory  shows that the vacuum energies obtained in the  world-sheet theory cannot be degenerate;
they are split in accordance with (\ref{ten}). Moreover, we will show
 below that with $|m_A-m_B|\neq 0$ and the $F$-term stabilization,
 \ntwoo supersymmetry in the world-sheet theory  
 spontaneously breaks already at the classical level.

\subsection{\boldmath{$|m_A-m_B|\neq 0$}}

Now we will construct the effective world-sheet theory for a 
non-Abelian string
in  \ntwo SQCD (\ref{model}) with $|m_A-m_B|\neq 0$ deformed by the mass 
term for the adjoint matter
(\ref{msuperpotbr}). There are two ways of addressing this problem. First, 
we can start
from  (\ref{model}) with  $\xi_3=0$ and   $|m_A-m_B| = 0$,
 see (\ref{equalmasses}). To the leading order in $\mu$ the deformation superpotential then
reduces to a single FI $F$-term -- that of the U(1) factor of the U$(N)$ gauge group,
with the complex FI parameter
\beq
\xi=\xi_1+i\xi_2=2\sqrt{\frac{2}{N}}\,\mu_1\, m\,,
\label{Fxi}
\eeq
see (\ref{xis}). In this limit the theory has unbroken \ntwo supersymmetry 
\cite{HSZ,VY},
as well as unbroken color-flavor symmetry (\ref{c+fxf}). In particular, 
the FI term
(\ref{Fxi}) can be rotated by a global SU(2)$_R$ transformation to the FI 
$D$-term $\xi_3$.
The world-sheet theory on the
non-Abelian string is given in this case by \ntwot CP model 
(\ref{wcp}), where all masses $m_A$ are equal.

Now we switch on the splittings $|m_A-m_B|\neq 0$ and ask 
ourselves: what is the response in the world-sheet 
theory? Clearly, the world-sheet theory becomes a certain
deformation of (\ref{wcp}), with generic mass parameters $m_A$.  
\ntwot supersymmetry breaking is expected. To see that this 
is indeed the case it is worth remembering  that the $Z_N$ strings under consideration
are BPS saturated only being considered in the
U(1)$^N$ Abelian theory, see Sect.~\ref{stringtensions}. 
In particular, we can introduce $N$ different FI $F$-terms
$\xi_P$ which determine central charges of $N$ different
BPS strings 
only in the Abelian U(1)$^N$ theory.

On the other hand, the non-Abelian
 strings we deal with
are, in fact, interpolations between different $Z_N$ strings 
\cite{HT1,ABEKY,SYmon,HT2}.
They exist only in the full non-Abelian U($N$) gauge theory. 
Therefore, non-Abelian strings
in the theory (\ref{model}) are {\em not} BPS saturated. To begin 
with, they all have different tensions, see (\ref{ten}). Since the 
world-sheet theory on
 the non-Abelian string describes dynamics of the
orientational modes  interpolating between different $Z_N$ strings,
we expect \ntwot world-sheet supersymmetry to be broken.

Another line of reasoning  is to start with a large nonvanishing $\xi_3$ and all 
$|m_A-m_B| = 0$.
Then one deforms the theory by adding  the superpotential (\ref{msuperpotbr}). 
 The world-sheet theory becomes 
 massless heterotic \ntwoo CP model.  Next, one introduces generic $m_A-m_B\neq 0$,
simultaneously  decreasing $\xi_3$ and  increasing $m$ keeping $\mu m$ large. Eventually one takes the limit
$\xi_3=0$, see (\ref{weakcoup}).

These two approaches combined, suggest that the world-sheet theory we are looking 
for is a heterotic
 CP model (\ref{wcp}),  with generic masses deformed by a certain
two-dimensional superpotential ${\mathcal W}_{1+1}(\sigma)$, which  
 breaks \ntwot supersymmetry down to \ntwoo . We assume this 
below.  To derive the deformation  superpotential we have to find, generally speaking,
solutions for non-Abelian strings and substitute them in the bulk 
action (\ref{model}) assuming a slow adiabatic dependence of the moduli $n^P$ 
and $\rho^K$ on the world-sheet coordinates, cf.  \cite{SYrev}.

We leave this program for future studies while for the time being we
 make a crucial shortcut. We determine the deformation potential
using the $Z_N$ string tensions   (\ref{ten}) as an input.
As was already mentioned,
the  $Z_N$ string tensions must be identified with 
the vacuum energies for $N$ vacua in the world-sheet theory,
\beq
T_P=E_{P}^{1+1}, \qquad P=1, ..., N\,,
\label{tenevac}
\eeq
where $E_{P}$ are the vacuum energies. For undeformed theory
 this identification  is usually carried out 
up to a constant shift, see \cite{SYrev}.
 Namely, the energies
of the $N$ vacua in \ntwot supersymmetric 
CP model (\ref{wcp}) all vanish, while the
$Z_N$ string tensions   are all equal to $2\pi\xi_3$.
In our case we identify vacuum energies of the world-sheet
theory with tensions of $Z_N$ strings without any shift.

 We use  relation  (\ref{tenevac}) below to find
the deformation potential $V_{1+1}(\sigma)$.
To this end we note that at small $\mu$ the potential $V_{1+1}(\sigma)$ is a small perturbation and to the leading 
order in $\mu$,  the  expectation values  $\sigma_P$ are (classically) given by their unperturbed values in 
Eq.~(\ref{classvac}). Since the vacuum energies for the undeformed \ntwot supersymmetric 
CP model (\ref{wcp})  all vanish, Eq.~(\ref{tenevac})
implies
\beq
T_P=V_{1+1}(\sigma_P), \qquad P=1,...,N,
\label{tenpot}
\eeq
where $\sigma_P$ are the expectation values  of 
the $\sigma$ field in $N$ vacua of the world-sheet theory.

Combining this with (\ref{ten}) and (\ref{classvac}) we determine the 
world-sheet potential, 
\beq
V_{1+1}(\sigma)=4\pi\left|\sqrt{\frac{2}{N}}\,\mu_1\,m-
\mu_2 \left(\sqrt{2}\sigma+m\right)\right|,
\label{2Dpot}
\eeq
where $m$ is the average of  masses of $N$ first quarks,
which condense in the  $r=N$ vacuum of the bulk theory, see
(\ref{avm}). 

The potential (\ref{2Dpot}) gives (to the leading order in $\mu$)
the vacuum energies for all $N$ vacua of the world-sheet theory right, i.e. equal
to tensions of the $N$ elementary strings. In principle, one could add
to (\ref{2Dpot}) an arbitrary potential, which vanishes in all
$N$ critical points (vacua). For now we assume   that 
this additional potential is zero. A rigorous proof of this
assertion will be presented in a future publication.

The deformation potential (\ref{2Dpot}) can be written as the
modulus squared of the derivative of a certain 
two-dimensional superpotential ${\mathcal W}_{1+1}(\sigma)$,
\beq
{\mathcal W}_{1+1} = \frac{1}{\mu_2}\,\sqrt{\frac{8\pi}{9}}\left[\sqrt{\frac{2}{N}}\,\mu_1\,m-
\mu_2 \left(\sqrt{2}\sigma+m\right)
\right]^{3/2}
\label{dopo}
\eeq
cf. (\ref{etong}). This confirms our initial assumption that 
the world-sheet theory has \ntwoo supersymmetry at the Lagrangian
level (broken spontaneously by the choice of vacua already at the classical level).

The vacuum energies of $N$ vacua in world-sheet theory are
nonvanishing and all different for a generic set of $m_A-m_B$. 
This shows that \ntwoo supersymmetry is broken at the
classical level.
 In 
\cite{BSYhetmass} we demonstrated, however, that the masses of 
the fermion and boson excitations
($n$'s, $\rho$'s vs. their fermion superpartners)
are still identical to the leading order in $\mu$. They
  split only at the next-to-leading order.
  
To conclude this section it is instructive to consider the 
limit of unbroken maximal supersymmetry. To this end we put
$\mu_2=0$ in the deformation superpotential of  the bulk theory (\ref{msuperpotbr}).  To the leading order in 
$\mu_1$ superpotential   (\ref{msuperpotbr}) reduces in this case to the single FI $F$-term of the U(1) factor of U$(N)$ gauge
group. This FI term does not break \ntwo supersymmetry (to the leading order in $\mu$),
and the  1/2 ``BPS-ness" of the string solution is maintained 
guaranteeing \ntwot supersymmetry on the world sheet.

In more detail,
with $\mu_2=0$ the deformation potential (\ref{2Dpot}) reduces to a  constant equal to the common value of 
the $Z_N$-string
tensions. This overall constant does not ruin the 
\ntwot supersymmetry of the weighted CP model 
(\ref{wcp}) on the string  (see the discussion in Sect.~6.3).

If $\mu_2\neq 0$ (and $\mu_2\sim \mu_1$), and the set of masses $m_P$ is generic (i.e. all $|m_P|$'s 
are of the same order of magnitude, none of these masses vanish
or are clustered in a special way) then the split of the string tensions is of the order of the
central value of the tension.\footnote{For simplicity we assume that $N$ does not grow. Otherwise, we should  take 
into account the $N$ dependence of $\Delta\,\sigma$.} It is instructive to compare this statement with the
$D$-term stabilized strings where the central value is proportional to $\xi$ (see \cite{SYhet})
while the split is proportional \cite{BSYhetmass} to $\mu^2m_P^2/\xi$ (in the limit of small deformation).
Given the identification $\xi\sim \mu m$, 
we conclude that the situations with the $F$- and $D$-term stabilized heterotically deformed strings
are qualitatively similar.

It is curious to mention a special case $m_P = m_0\exp\left(\frac{2\pi\, i \, P}{N}\right)$ where $P=1,2,..., N$. 
In this case $m$ vanishes, and Eq. (\ref{2Dpot}) reduces to $4\sqrt{2}\pi \,|\mu_2\, \sigma |$
(supplemented by (\ref{sigmaeq}) implying for large $m_P$ that $E_P = 4 \pi \,|\mu_2\, m_0|$).
All tensions are the same. This is due to the $Z_N$ symmetry of this example.

To conclude this section, it is worth summarizing our findings regarding the pattern of supersymmetry
breaking in the world-sheet theory.
Generally speaking, there are four mass parameters in the problem at hand,
$m$ (the average squark mass term), $\Delta m$ (or, alternatively, $\Delta m_{AB}$,
typical squark mass differences),  $\Lambda$
(the dynamical scale parameter), and -- finally -- $\mu$ (the deformation
parameter).
For simplicity let us assume $\Lambda$ to be very small and negligible.
Then we are left with 3 parameters.
In the limit $\mu \to 0, \,\, m \to  \infty,\,\, (\mu m) \,\, {\rm fixed}$, {\em and}
$$\Delta m=0$$
the bulk theory has ${\mathcal N}=  2$,
the strings are BPS-saturated, and the theory on the world sheet possesses
${\mathcal N}= (2,2)$ too  (to the leading order in $\mu$).
All $N$ strings are degenerate. The tension scales as $\mu m$.
(The first string studies in this limit were carried out by Bolognesi \cite{stefano}.)

It is worth explicitly verifying ${\mathcal N}= (2,2)$ supersymmetry in the world-sheet theory at
$\Delta m_{AB}=0$. The field $\sigma$ is not dynamical and can be eliminated
by virtue of  its equations of motion. In the case $\Delta m_{AB}=0$ we have then
$\sqrt{2}\sigma=-m$.
Thus, the second term  in the potential (\ref{2Dpot}) vanishes, and the potential
reduces to a constant which gives  common tension to all $Z_N$ strings.
This overall constant does {\em not}   break ${\mathcal N}= (2,2)$ 
supersymmetry on the world sheet (see Sect.~6.3).

Now, if we switch on $\Delta m \neq 0$, a superpotential is generated
on the world sheet which, generally speaking,
breaks the world-sheet supersymmetry down to ${\mathcal N}= (0,2)$ as far as algebra is
concerned (i.e. in the Lagrangian).
This  world-sheet ${\mathcal N}= (0,2)$ is further
{\em spontaneously}  broken down to nothing.

Let us stress that if we consider the next-to-leading order in $\mu$, the world-sheet 
supersymmetry will be explicitly and completely broken. At $O(\mu^2)$ we expect generation of the
potential which cannot be presented  as the modulus squared  (of the derivative) of a certain
superpotential. One can  expect that this potential may depend
directly on the $n$ fields in addition to the   $\sigma$ dependence.

\subsection{Linear \boldmath${\mathcal W}_{1+1} (\sigma)$}}
\label{linsupepop}

Formally, the insertion of an ``additional"
 superpotential ${\mathcal W}_{1+1} (\sigma)$ in the weighted \ntwot CP model (which can be obtained
 as a dimensional reduction of a super-QED from four to two dimensions)
 implies that \ntwot supersym\-metry 
 on the world sheet is explicitly broken down to \ntwoo which, in turn,   may or may not
be spontaneously broken down to nothing. In fact, there is one exception from this rule.
Indeed, assume ${\mathcal W}_{1+1} (\sigma)$ to be a linear function of $\sigma$. As is obvious e.g.
from Eqs.  (4.1) and (4.2) of \cite{SYhet}, the part of the heterotically deformed Lagrangian which 
describes interactions contains only  $\partial^2{\mathcal W}_{1+1}/\partial \sigma^2$. For linear superpotentials 
it vanishes, and we have exactly the same Lagrangian as that of the \ntwot model, up to an overall 
constant shift of energy
proportional to $|\partial{\mathcal W}_{1+1}/\partial \sigma |^2$.
If we remove this constant shift of the vacuum energy by hand, the remainder
satisfies the \ntwot superalgebra. In other words, the linear deformation superpotential leads to
$E_0 +{\mathcal L}_{{\mathcal N}=(2,2)}\,$, where $E_0$ is a numerical 
constant.\footnote{$E_0$ can be viewed as a central charge in the \ntwot superalgebra.}
This phenomenon is somewhat similar to a well-known fact in four dimensions. 
If we consider \ntwo gauge theories, then a generic superpotential
${\mathcal W}({\mathcal A})\neq 0$ breaks \ntwo down to \none
(here ${\mathcal A}$ is the \ntwo  photon/photino superpartner). However, linear ${\mathcal W}({\mathcal A})$
does preserve \cite{HSZ,VY} the full  \ntwo supersymmetry. This is   an exception too.

\vspace{4mm}

\section{Quantum effects in the world-sheet theory}
\label{2Dquant}
\setcounter{equation}{0}

In this section we study quantum (nonperturbative) effects in the world-sheet theory and show that
quantum corrections to the vacuum energies of $N$ vacua precisely reproduce
quantum corrections to string tensions (\ref{qten}) obtained in the bulk theory.
We start from reviewing the exact superpotential in the undeformed \ntwot CP model (\ref{wcp})
and then switch on the deformation potential (\ref{2Dpot}).

\subsection{Exact superpotential}
\label{secsup}

The \ntwot supersymmetric \cpn models are known to be described by an exact superpotential \cite{AdDVecSal,ChVa,W93,Dorey} of
the  Veneziano--Yankielowicz  type \cite{VYan}.
This superpotential was generalized to the case of the weighted CP models in
\cite{HaHo,DoHoTo}. In this section we will briefly outline  this method.
Integrating out the fields $n^P$ and $\rho^K$  we can describe
 the original model  (\ref{wcp}) by the following
exact twisted superpotential:
\beqn
 {\cal W}_{\rm eff}
 & =&
\frac1{4\pi}\sum_{P=1}^N\,
\left(\sqrt{2}\,\Sigma+{m}_P\right)
\,\ln{\frac{\sqrt{2}\,\Sigma+{m}_P}{\Lambda}}
\nonumber\\[3mm]
&-&
\frac1{4\pi}\sum_{K=N+1}^{N_F}\,
\left(\sqrt{2}\,\Sigma+{m}_K\right)
\,\ln{\frac{\sqrt{2}\,\Sigma+{m}_K}{\Lambda}}
\nonumber\\[3mm]
&-& \frac{N-\tN}{4\pi} \,\sqrt{2}\,\Sigma\, ,
\label{2Dsup}
\eeqn
where $\Sigma$ is a twisted superfield \cite{W93} (with $\sigma$ being its lowest scalar
component).
Minimizing this superpotential with
respect to $\sigma$ we get the vacuum field formula,
\beq
\prod_{P=1}^N(\sqrt{2}\,\sigma+{m}_P)
=\Lambda^{(N-\tN)}\,\prod_{K=N+1}^{N_f}(\sqrt{2}\,\sigma+{m}_K)\,.
\label{sigmaeq}
\eeq

Note, that the roots of this equation coincide with the double roots of the Seiberg--Witten curve  of
the bulk theory \cite{Dorey,DoHoTo} for all $N_f<2N-1$.
Below  we will see that this fact is crucial
for the relation between the tensions of 
the non-Abelian strings and vacuum energies of the world-sheet
theories in the $\mu$-deformed quantum theory.
This coincidence is, of course, a manifestation of the coincidence
of the Seiberg--Witten solution of the bulk theory in the $r=N$ vacuum 
with the exact solution of the two-dimensional model (\ref{wcp})
defined by the superpotential (\ref{2Dsup}). As was mentioned in Sec.~\ref{intro},
this coincidence was observed in \cite{Dorey,DoHoTo,DTo} and explained later
in \cite{SYmon,HT2}.

In particular, in the example $N_f=N=2$ considered in Sect.~\ref{quantum} and Appendix A
the vacuum equation (\ref{sigmaeq}) reduces to
\beq
(\sqrt{2}\,\sigma+{m}_1)(\sqrt{2}\,\sigma+{m}_2)
=\Lambda^{2}\,.
\label{sigmaeq20}
\eeq
It has two solutions
\beq
\sqrt{2}\sigma_{1,2}=-m \mp \sqrt{\frac{\Delta m_{12}^2}{4} +\Lambda^2},
\label{sigma20}
\eeq
which indeed coincide with double roots (\ref{eN2}) of the Seiberg--Witten curve,
\beq
\sigma_P=e_P, \qquad N_f<2N-1.
\label{sigmae}
\eeq
Another example (considered in Appendix B) is the one with $N=3$ and $N_f=5$.
This is a special  example of the case $N_f=2N-1$. Restricting ourselves to the mass choice
(\ref{massch}) we easily find VEVs of $\sigma$ for this case from (\ref{sigmaeq}), namely
\beq
\sqrt{2}\sigma_1=-m_1, \qquad \sqrt{2}\sigma_2=-m_2, \qquad \sqrt{2}\sigma_3=-m_3
 +\Lambda\,.
\label{sigma32}
\eeq
We see that in this case the relation between VEVs of $\sigma$ and double roots of the
Seiberg--Witten curve is modified, namely,
\beq
\sqrt{2}\sigma_P=\sqrt{2}e_P+\frac{\Lambda}{3}, \qquad N=3, \quad N_f=5,
\label{sigmae32}
\eeq
see (\ref{eN3}).
If $N$ is an arbitrary integer (rather than $N=3$) this relation takes the form
\beq
\sqrt{2}\sigma_P=\sqrt{2}e_P+\frac{\Lambda}{N}, \qquad N_f=2N-1,
\label{sigmaespecial}
\eeq
see (\ref{shift}).

\subsection{Quantum corrections to vacuum energies}
\label{qctve}

Now let us demonstrate that quantum (nonperturbative) corrections to the
vacuum energies
in the world-sheet theory precisely reproduce the quantum-corrected string
tensions (\ref{qten}). The vacuum energies in the $N$ vacua
of the world-sheet theory are given by the values of the 
world-sheet deformation potential $V_{1+1}(\sigma)$
calculated at VEVs of the $\sigma$ field. Much in the same way
as in the classical theory, the VEVs of the $\sigma$ field are determined
by the undeformed \ntwot supersymmetric theory (\ref{wcp})
to the leading order in $\mu$. 

In quantum theory these VEVs -- we denote them as $\sigma_P$ --  are presented by 
solutions of the vacuum equation (\ref{sigmaeq}). Consider
the case $N_f<2N-1$ first. In this case the potential (\ref{2Dpot}) gives
\beq
T^{\rm BPS}_P=V_{1+1}(\sigma_P)=4\pi\,\left|\sqrt{\frac{2}{N}}\,\mu_1\,m-\mu_2\left(\sqrt{2}\sigma_P+m\right)\right|.
\label{qenergies}
\eeq
This formula exactly reproduces our result (\ref{qten}) 
obtained in the bulk theory, provided that the
relations (\ref{sigmae}) and (\ref{em}) are taken into account.

As far as the special case $N_f=2N-1$ is concerned, the 
deformation potential (\ref{2Dpot}) should be modified.
The modification is 
\beq
V_{1+1}(\sigma)=4\pi\left|\sqrt{\frac{2}{N}}\,\mu_1\,m-
\mu_2\left(\sqrt{2}\sigma-\frac{\Lambda}{N}+m\right)\right|,
\quad N_f=2N-1\,.
\label{2Dpotspec}
\eeq
This potential reproduces  the string tensions (\ref{qten})
once we make use of the relation (\ref{sigmaespecial}).
The superpotential (\ref{dopo}) is modified accordingly.

\section{Monopole confinement}
\label{monopole}
\setcounter{equation}{0}

Now we will discuss kinks on the $F$-term stabilized
strings, which represent confined monopoles
being viewed from the bulk standpoint.

Consider weak coupling regime (\ref{weakcoup}) in the 
bulk theory. Since $N$ squarks are   condensed in the
$r=N$ vacuum (see (\ref{qvev})) and the gauge group is Higgsed
the 't Hooft-Polyakov mono\-poles
are confined. As we know, in the Higgsed U$(N)$ gauge theories monopoles show up 
only as junctions of two distinct elementary non-Abelian strings \cite{Tong,SYmon,HT2}.  
These strings  are in fact represented by different  vacua in the effective
world-sheet  sigma model   while the
 confined monopoles are   kinks interpolating
between distinct vacua \cite{SYmon,HT2,Tong}.

Given  the bulk theory (\ref{model}) let us inspect the domain
of small $|\Delta m_{AB}|\ll m$, with $m$
  large enough to ensure (\ref{weakcoup}) at small $\mu$.
In this case the splittings between different vacua of 
the world-sheet theory are small. We can consider them as 
``quasivacua."

This regime is quite similar to the one  studied in \cite{GSY05} in non-supersymmet\-ric
bulk theory in the Higgs phase, where all quark condensates   are equal. In the effective  
\cpn model on the non-Abelian string all vacua  are  split (on the quantum level),  and $N-1$ would-be
vacua become quasivacua, see \cite{SYrev}
for a review. The vacuum splitting can be understood as a manifestation of the Coulomb/confining linear
potential between the kinks \cite{Coleman,W79} that interpolate between the true vacuum, and say,
the lowest quasivacuum. The force is attractive in the kink-antikink pairs, implying  formation
of weakly coupled bound states (weak coupling is the
manifestation of the smallness of the splittings between the vacua).
The charged kinks are eliminated from the spectrum, see 
Fig.~\ref{fig:conf}.

\begin{figure}
\epsfxsize=8cm
\centerline{\epsfbox{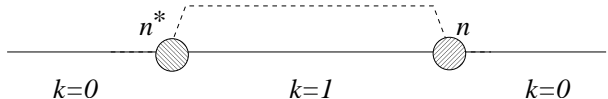}}
\caption{\small 
Linear confinement of the kink-antikink pair.
The solid straight line represents the ground state.
The dashed line shows
the vacuum energy density of the lowest quasivacuum.}
\label{fig:conf}
\end{figure}

The
kink confinement in the two-dimensional CP model can be interpreted \cite{GSY05} as 
the following phenomenon:
the non-Abelian monopoles,
in addition to  the four-dimensional confinement
(which ensures that the monopoles
are attached to the strings) acquire a two-dimensional confinement along
the string:
a monopole--antimo\-nopole   forms a meson-like configuration,
with necessity, see Fig.~\ref{fig:conf}.

Moreover,  as was shown in \cite{HoVa} for the \cpn model 
and in \cite{SYtorkink} for the weighted CP model (\ref{wcp}),
the kinks belong to the fundamental 
$(N,1)+(1,\tN)$ representations of the global group (\ref{c+fxf}) in the  limit (\ref{equalmasses}).
The global flavor group is explicitly broken by the mass differences,  down to U(1)$^{N_f-1}$.
The kinks are charged with respect to this group. Therefore,
the kink-antikink  (monopole-antimonopole) mesons which carry non-trivial charges with respect to 
the global U(1)$^{N_f-1}$
 are stable. In other words, if the total global charge of such a 
 meson does not vanish, the  kink and antikink cannot annihilate, and the meson
 they constitute will never decay.
 
\section{Generic single-trace superpotential \\
deformations}
\label{genericsup}
\setcounter{equation}{0} 

In this section we will treat  more general  deformation of the bulk theory than those considered previously. 
Namely, instead of (\ref{msuperpotbr}) we will consider  a   generic single-trace polynomial
superpotential of the form 
\beq
\label{brsup}
{\mathcal W}_{3+1}={\rm Tr}\,\sum_{k=1}^{N}\frac{c_k}{k+1}\,\Phi^{k+1}\,,
\eeq
where we introduce the adjoint matrix superfield
\beq
\Phi=\frac12 \, {\mathcal A}+ T^a {\mathcal A}^a\,.
\label{admatrix}
\eeq
$\Phi$ in Eq. (\ref{admatrix})  is a matrix from U($N$) rather than SU($N$).
 The bosonic potential of the bulk theory is still 
given by (\ref{pot}). We will require the coefficients $c_k$
to be  small and study the theory's response  in the leading approximation in
$c_k$. This condition is quite similar to the condition  (\ref{smallmu}) of 
small $\mu$.

Such more generic deformations were considered in \cite{Edalati,SYhet}
with  a special choice of superpotentials:
their critical points we supposed to coincide
with the quark mass terms. Here we {\em relax} this condition and, instead,
put $\xi_3=0$, much in the same way as for the deformation (\ref{msuperpotbr}),
so the quark condensation is entirely due to the
FI $F$-terms.  The gauge group U($N$) is
broken down to U(1)$^N$ by the adjoint  VEVs (\ref{avev}).

The squark VEVs can be readily
calculated from (\ref{pot}). They are still given by Eq.~(\ref{qvev}), where now
\beq
(\xi_1,...,\xi_N)=-{\rm diag}\left\{\sqrt{2}\,\, \frac{\pt{\mathcal W}_{3+1}}{\pt \Phi}(\Phi_{\,\rm VEV})\right\}.
\label{xiW}
\eeq
Here $\Phi_{\,\rm VEV}$ is the vacuum expectation value of the
matrix field  $\Phi$ defined in (\ref{avev}).

Consider now the $Z_N$ strings in this theory. If the deformation superpotential (\ref{brsup}) is weak,
 these strings are
BPS saturated  in the low-energy U(1)$^{N}$ gauge theory.
Parallelizing  the derivation presented in Sec.~\ref{stringtensions} we get the same
expression (\ref{ten}) for their tensions, where $\xi_P$
now are given in Eq.~(\ref{xiW}).

Next we consider the response of 
the world-sheet theory on the non-Abelian string
on the bulk deformation
(\ref{brsup}). The world-sheet theory is still given by the weighted
CP model (\ref{wcp}) deformed by a certain two-dimensional
superpotential ${\mathcal W}_{1+1}(\sigma)$, which
breaks \ntwot supersymmetry down to \ntwoo\!. 

To find this deformation superpotential
we again impose the condition (\ref{tenpot}). Since the deformation is weak, by assumption,
the  $\sigma$ field VEVs   can be determined 
in the undeformed theory (\ref{wcp}).
They are still (classically) given by quark mass parameters via
(\ref{classvac}). This leads us to the following deformation potential
\beq
V_{1+1}(\sigma)=2\pi\sqrt{2}\,\left|\frac{\pt{\mathcal W}_{3+1}(\sigma)}{\pt \sigma}\right|\ .
\label{2DpotW}
\eeq
Note, that this potential still can be written in 
terms of certain
two dimensional superpotential ${\mathcal W}_{1+1}(\sigma)$
via identification (\ref{etong}),
\beq
{\mathcal W}_{1+1} = \sqrt{2\pi\sqrt{2}}\,\int^{\sigma} \, d\sigma^{'} \,\sqrt{\frac{\pt{\mathcal W}_{3+1}}{\pt \sigma^{'}}}\,.
\label{2ddsp}
\eeq
This fact shows the presence
of \ntwoo supersymmetry in the world-sheet theory at the
Lagrangian level, as was expected.

The simplest example is the quadratic superpotential
\beq
\label{quadsup}
{\mathcal W}_{3+1}=\mu_2\,{\rm Tr}\,\Phi^{2}\,,
\eeq
which can be compared to the adjoint mass deformation 
(\ref{msuperpotbr}) we studied before. For the superpotential
(\ref{quadsup}) Eq.~(\ref{2DpotW}) gives
\beq
V_{1+1}(\sigma)=4\pi\sqrt{2}\,\left|\mu_2\,\sigma\right|.
\label{2DpotWqud}
\eeq
In fact, (\ref{msuperpotbr}) is not a single trace superpotential for generic $\mu_1$ 
and $\mu_2$. However, if we take 
\beq
\mu_1=\sqrt{\frac{N}{2}}\,\mu_2\,,
\label{singletrace}
\eeq
the superpotential (\ref{msuperpotbr}) becomes equal to
(\ref{quadsup}). It is worth observing that with this choice of
$\mu_1$, the potential (\ref{2DpotWqud}) coincides with that
in Eq.~(\ref{2Dpot}).

The results we obtained in Sect.~\ref{2Dquant} suggest that 
the potential (\ref{2DpotW}) correctly reproduces
the quantum  corrections  $O(\Lambda/m_A)$ to 
the classical string tensions. Namely, to reproduce the
quantum corrections to
(\ref{ten}) with $\xi$'s given in Eq.~(\ref{xiW}) we  calculate
 (\ref{2DpotW}) at the critical points $\sigma_P$. The latter are
given by the
solutions of Eq.~(\ref{sigmaeq}) (for $N_f<2N-1$).

In the special case $N_f=2N-1$ we expect that, 
in order to reproduce the bulk quantum corrections, Eq.~(\ref{2DpotW})
should be modified as follows:
\beq
V_{1+1}(\sigma)=2\pi\sqrt{2}\,\left|
\frac{\pt{\mathcal W}_{3+1}}
{\pt \sigma}\left(\sigma-\frac{\Lambda}{\sqrt{2}\,N}
\right)\right|,
\qquad N_f=2N-1,
\label{2DpotWspec}
\eeq
 see (\ref{2Dpotspec}).

\section{Conclusions}
\label{concl}
\setcounter{equation}{0}

We studied heterotic deformations in the problem of the $F$-term stabilized non-Abelian
strings. The bulk theory supporting these strings is
\ntwo SQCD with no Fayet--Iliopoulos $D$-term in which \ntwo supersymmetry is broken
down to \none
by superpotentials of the type (2.2) (or generic higher order polynomials).
In the limit of weak deformation we found the heterotic superpotential appearing in the
weighted CP model on the string world sheet which breaks \ntwot supersymmetry down
to \ntwoo. The latter world-sheet supersymmetry is further  spontaneously broken at the tree level,
generally speaking.
Our results dramatically expand the class of heterotic models which are generated on
the non-Abelian strings
in the \none bulk theories.

The potential (6.6) linear in $\mu$ is the leading order-potential. Say,
in the case of equality
of all squark masses  it gives a common tension to all $N$ strings.
The  world-sheet supersymmetry is \ntwot. The next-to leading order terms in the
potential are $O(\mu^2)$.
They break   supersymmetry completely. Our finding is that if we restrict
ourselves to
the leading (linear) order in $\mu$, but consider nondegenerate masses the
world-sheet supersymmetry is explicitly  broken down  to \ntwoo.

\section*{Acknowledgments}

The work of MS was supported in part by DOE grant DE-FG02-94ER408. 
 The work of AY was  supported
by  FTPI, University of Minnesota,
by RFBR Grant No. 09-02-00457a
and by Russian State Grant for
Scientific Schools RSGSS-11242003.2.

\section*{Appendix A: 
U(2) with \boldmath{$N_f=2$}}

 \renewcommand{\theequation}{A.\arabic{equation}}
\setcounter{equation}{0}

 \renewcommand{\thesubsection}{A.\arabic{subsection}}
\setcounter{subsection}{0}

In this Appendix we consider the 
simplest example: U(2) gauge theory with $N_f=2$ flavors. We calculate
the diagonal elements of the matrix $E$, see Eq.~(\ref{E}), given by
\beq
E=\frac1{2}\,\frac{\pt u_2}{\pt a}+\frac{\tau^3}{2}\,\frac{\pt u_2}{\pt a^3}
\label{EN2}
\eeq
in this particular case.

The exact solution of the theory on the Coulomb branch relates 
the fields $a$ and $a^3$ to 
contour integrals running along the contours $\alpha_i$ ($i=1,2$) encircling  the double roots $e_1$ and $e_2$
(in the anticlockwise direction), see (\ref{rNcurve}) for $N=2$.
Using explicit expressions from \cite{ArFa,KLTY,ArPlSh,HaOz} and generalizing them to 
the U($N$) case\,\footnote{This amounts to including derivatives with respect to $u_1$ and   terms proportional to the average $e$ below (or $m$, which is related to $e$ via (\ref{em})). These terms are absent in
SU$(N)$ case considered in \cite{ArFa,KLTY,ArPlSh,HaOz}.}
we write
\beq
\frac{\pt \Phi^i}{\pt u_2}= \frac12 \,\frac1{2\pi i}\oint_{\alpha_i}  \frac{dx}{y}\,,\,\,
\qquad \frac{\pt \Phi^i}{\pt u_1}=  \,\frac1{2\pi i}\oint_{\alpha_i} \frac{dx}{y}
\left[x-(e_1+e_2)\right]\,,
\label{dadu20}
\eeq
where we define
\beq
(\Phi_1,,...,\Phi_N)={\rm diag}\left(\frac12\, a + T^{\tilde{a}}\, a^{\tilde{a}}\right),
\label{Phi}
\eeq
which gives in the $N=2$ case 
\beq
a = \Phi_1+\Phi_2, \qquad a^3=\Phi_1-\Phi_2.
\label{PhiN2}
\eeq

For the factorized curve (\ref{rNcurve}) the integrals (\ref{dadu20}) can be  easily
calculated and are given by their pole contributions. This provides us with the derivatives
$\pt a/\pt u_1$, $\pt a^3/\pt u_1$, $\pt a/\pt u_2$ and $\pt a^3/\pt u_2$.
Inverting this matrix we obtain the desired derivatives
\beq
\frac{\pt u_2}{\pt a}=e_1+e_2, \qquad \frac{\pt u_2}{\pt a^3}=e_1-e_2\,.
\label{duda20}
\eeq
Finally, substituting this in (\ref{EN2}) we get
\beq
{\rm diag}\,E=(e_1,e_2)\,.
\label{EN2e}
\eeq
Then the result for the string tensions (\ref{qten}) ensues.
Now $e_1$ and $e_2$
are the two double roots of the curve.
For the sake of completeness we present their explicit form (see, for example, \cite{SYcross}),
\beq
\sqrt{2}\,
e_{1,2}=-m \mp \sqrt{\frac{\Delta m_{12}^2}{4} +\Lambda^2}\,.
\label{eN2a}
\eeq
Substituting (\ref{eN2a}) in (\ref{qten}) we get tensions of two $Z_2$ strings in
(\ref{qten20}).

As was already mentioned, we can relax the weak-coupling conditions (\ref{U1N}) and
(\ref{weakcoup}) and go to the strong-coupling domain (\ref{strcoup}). The theory undergoes
a crossover transition, i.e. crosses curves of marginal stability (CMS), where certain states decay.
Also, we pass through monodromies
which change charges of other states. In this domain one can exploit a dual
 description. The dual to the original theory is weakly coupled U(1)$^2$ gauge theory
of light dyons,
see \cite{SYcross} for details. Two U(1) gauge fields interacting with the light dyons are now
$A_{\mu}$ and $B_{\mu}=1/\sqrt{5}(A_{\mu}^3+2A_{\mu}^{3D})$, where $A_{\mu}^{3D}$ is
the dual gauge potential. Their scalar superpartners are
\beq
a, \qquad b=\frac1{\sqrt{5}}(a^3+2a^{3}_D)\, .
\label{ab}
\eeq
The charges of two light dyons with respect to these U(1) fields are \cite{SYcross}
\beq
\left(\frac12,\,\,\pm \frac{\sqrt{5}}{2}\right).
\eeq
Repeating the same steps which led us to (\ref{qten}) in the dual theory we
get
\beq
T ^{\rm BPS}_{1,2}=4\pi \sqrt{2}\,\left|\, \mu_1\,\frac1N\,\frac{\pt u_2}{\pt a} \pm \frac{\mu_2}{2\sqrt{5}}
\frac{\pt u_2}{\pt b}\,\right|.
\label{qtendual}
\eeq
The monodromies mentioned above change the
root pairing   but does not change the factorized form of the curve
(\ref{rNcurve}). Therefore, passing to the dual theory we have
\beq
\frac{\pt u_2}{\pt a}\to \frac{\pt u_2}{\pt a}=e_1+e_2, \qquad
\frac{\pt u_2}{\pt a^3} \to \frac{1}{\sqrt{5}}\frac{\pt u_2}{\pt b} =e_1-e_2\,,
\label{atob}
\eeq
which gives us the same expressions (\ref{qten}) and (\ref{qten20}) for the string tensions.

We see that our result (\ref{qten}) for the $Z_N$ string tensions  can be analytically
continued (as functions of $\Delta m_{AB}$ and $m$) in the strong-coupling domain.

\section*{Appendix B:  U(3) with \boldmath{$N_f=5$}}

 \renewcommand{\theequation}{B.\arabic{equation}}
\setcounter{equation}{0}

Let us now consider another example pertinent to  $N_f>N$, namely, the U(3) gauge theory with $N_f=5$. 
The matrix
$E$ now has the form
\beq
E=\frac1{2}\,\frac{\pt u_2}{\pt a}+T^3\,\frac{\pt u_2}{\pt a^3} +T^8\,\frac{\pt u_2}{\pt a^8}\,,
\label{EN3}
\eeq
where
\beq
T^3 =  \frac1{2}
 \left(
\begin{array}{ccc}
1 & 0 & 0 \\
0 & -1 & 0\\
0 & 0 & 0\\
\end{array}
\right), \qquad
T^8 =  \frac1{2\sqrt{3}}
 \left(
\begin{array}{ccc}
1 & 0 & 0 \\
0 & 1 & 0\\
0 & 0 & -2\\
\end{array}
\right).
\label{T3T8}
\eeq
To determine the derivatives of $u_2$ with respect to $a$, $a_3$ and $a_8$ we use the exact solution
of the theory on the Coulomb branch. Generalizing the SU(3) solution \cite{ArFa,KLTY,ArPlSh,HaOz}
 to the U(3) case we can write
\beqn
\frac{\pt \Phi^i}{\pt u_3}
&= &
 \frac13\,\frac1{2\pi i}\oint_{\alpha_i} \frac{dx}{y}\,,
\nonumber\\[3mm]
\frac{\pt \Phi^i}{\pt u_2}
&=&
 \frac12 \,\frac1{2\pi i}\oint_{\alpha_i}  \frac{dx}{y}\left[x-\frac13\left(e_1+e_2+e_3\right)\right],
\nonumber\\[3mm]
\frac{\pt \Phi^i}{\pt u_1}
&=&
 -\frac1{\pi i}\oint_{\alpha_i}  \frac{dx}{y}\left[x^2-\frac12 x\left(e_1+e_2+e_3\right)
\right.
\nonumber\\[3mm]
&+&
\left.
\frac19 \left(e_1+e_2+e_3\right)^2\right] +1\,.
\label{dadu32}
\eeqn
The above integrals  are readily
calculable  in the case of the  factorized Seiberg--Witten curve (\ref{rNcurve}). This gives us the derivatives
of $a$, $a^3$ and $a^8$ with respect to $u_1$, $u_2$ and $u_3$. Inverting the matrix of these derivatives is an
algebraic albeit  tedious calculation. Omitting all details we present the final answer,
\beqn
\frac{\pt u_2}{\pt a}
&=&
e_1+e_2+e_3\,, \qquad \frac{\pt u_2}{\pt a^3}=e_1-e_2\,,
\nonumber\\[3mm]
 \frac{\pt u_2}{\pt a^8}
 &=&
 \frac1{\sqrt{3}}(e_1+e_2-2e_3)\,.
\label{duda32}
\eeqn
Substituting this in Eq.~(\ref{EN3}) we get
\beq
{\rm diag}\,E=(e_1,e_2,e_3)\,,
\label{EN3e}
\eeq
which leads us to the final expression (\ref{qten}) for the $Z_3$ string tensions,
where now $e_P$'s  are the double roots of the Seiberg--Witten curve (\ref{rNcurve}).

It is instructive to give more explicit expressions for the string tensions in terms of $m_A$.
To this end we consider
a particular mass choice,
\beq
m_1=m_4, \qquad m_2=m_5.
\label{massch}
\eeq
With this choice, the roots of the curve can be  easily found \cite{SYdual},
\beq
\sqrt{2}e_1=-\tilde{m}_1\,, \qquad \sqrt{2}e_2=-\tilde{m}_2\,, \qquad \sqrt{2}e_3=-\tilde{m}_3
 +\Lambda \,,
\label{eN3}
\eeq
where $\tilde{m}_{A}$ are defined in (\ref{shift}) for the special case $N_f=2N-1$.
Substituting this in (\ref{qten}) we get the tensions of the three $Z_3$ strings,
\beqn
T ^{BPS}_{1}
&=&
4\pi\,\left|\sqrt{\frac23}\mu_1\,m + \mu_2 \left(m_1-m +\frac{\Lambda}{3}\right) \right|,
\nonumber\\
T ^{BPS}_{2}
&=&
4\pi\,\left|\sqrt{\frac23}\mu_1\,m + \mu_2 \left(m_2-m +\frac{\Lambda}{3}\right)\right|,
\nonumber\\
T ^{BPS}_{3}
&=&
4\pi\,\left|\sqrt{\frac23}\mu_1\,m + \mu_2 \left(m_3-m -\frac{2}{3}\Lambda\right)\right| .
\label{qten32}
\eeqn
Note that the relation (\ref{em}) is still fulfilled due to the mass shifts in
(\ref{shift}) which must be done if $N_f=2N-1$.

Similarly to the $N=N_f=2$ case we can continue our theory to the strong-coupling domain
(\ref{strcoup}), where it is described by the non-Abelian  theory
with the dual gauge group U(2)$\times$ U(1) and $N_f=5$ flavors of light dyons \cite{SYdual}.
The latter is not asymptotically free.
Using the low-energy effective action of this theory found previously  in \cite{SYdual} it is straightforward
to demonstrate that, although the light state charges   are changed due to monodromies in
the strong-coupling domain, the resulting expressions (\ref{qten32}) for the string tensions
experience no   change.
Thus, the string tensions have analytic behavior in $m_A$ across the crossover lines,
much in the same as as in the example we dealt with in Appendix A.

\end{document}